\documentclass[%
 aip,
 amsmath,amssymb,
 reprint,%
]{revtex4-2}

\usepackage{graphicx}% Include figure files
\usepackage{dcolumn}% Align table columns on decimal point
\usepackage{bm}% bold math
\usepackage[utf8]{inputenc}
\usepackage[T1]{fontenc}
\usepackage{mathptmx}
\usepackage{mathtools}
\usepackage{etoolbox}
\usepackage{xcolor}

\usepackage{twoopt}
\usepackage{mathtools}
\usepackage{orcidlink}

\newcommand{\rem}[1]{ } % Comment text out
\newcommand{\newtext}[1]{#1}

\makeatletter
\def\@email#1#2{%
 \endgroup
 \patchcmd{\titleblock@produce}
  {\frontmatter@RRAPformat}
  {\frontmatter@RRAPformat{\produce@RRAP{*#1\href{mailto:#2}{#2}}}\frontmatter@RRAPformat}
  {}{}
}%
\makeatother

\begin{document}

\title{Eigenmodes in an ultra-relativistic ultra-magnetized pair QED-plasma}

%\author[0000-0001-5987-2856]{Mikhail Medvedev}
%\author{Mikhail Medvedev}
%\affiliation{Department of Physics and Astronomy, University of Kansas, Lawrence, KS 66045, USA}
%\affiliation{Laboratory for Nuclear Science, Massachusetts Institute of Technology, Cambridge, MA 02139, USA}

\author{\orcidlink{0000-0002-5024-0075}Ryan T. Low}
\email{rtlow@ku.edu}
\affiliation{Department of Physics and Astronomy, University of Kansas, Lawrence, KS 66045}
\author{\orcidlink{0000-0001-5987-2856}{Mikhail V. Medvedev}}
%\email{medvedev@ku.edu}
\affiliation{Department of Physics and Astronomy, University of Kansas, Lawrence, KS 66045}
\affiliation{Laboratory for Nuclear Science, Massachusetts Institute of Technology, Cambridge, MA 02139}

\begin{abstract}
Ultra-relativistic quantum-electrodynamic (QED) plasmas, characterized by magnetic field strengths approaching and  exceeding the Schwinger field of approximately $B_{Q} \approx 4 \times 10^{13}$ gauss, hold significant interest for laser-plasma experiments and astrophysical observations of neutron stars and magnetars.
\newtext{A previous study has developed the QED-plasma framework, a framework for analyzing the collective plasma phenomena for arbitrarily strong magnetic fields, especially in the ultra-magnetized super-Schwinger-field regime.
In that study, QED modifications to plasma normal modes characterized by the linear regime in a cold plasma have been explored.} In the current study, we extend the previous analysis to
investigate the joint modification of normal plasma modes in ultra-relativistic electron-positron plasmas, both charge neutral and non-neutral, by the super-strong magnetic field and plasma relativistic temperature. Our analysis shows that the most substantial modification concerns the reduction of the plasma frequency cutoff, resulting in relativistic and field-induced transparency. Additionally, we observe a temperature-independent modification of the index of refraction of electromagnetic waves, which coincides with the behavior observed in a cold QED plasma. 
\end{abstract}


%\keywords{Plasma astrophysics -- Magnetic fields}


\maketitle

\section{Introduction}

The study of plasma behavior in ultra-strong magnetic fields, particularly those approcung or even exceeding the Schwinger field strength $B_Q = {m_e^2c^2}/{e\hbar} \sim 4 \times 10^{13}\textrm{ gauss}$, represents a novel frontier in modern plasma research.  

Existing and planned laser facilities, such as the Ultrashort Pulse Laser System (ZEUS), the Center for Relativistic Laser Science (CORELS), and the Laser Und XFEL Experiment (LUXE), are attempting to achieve field strengths and particle energies at which the quantum-electrodynamic (QED) effects become significant \citep{Bromage+19, Danson+19, Borysova+21}. However, the laser facilities primarily explore conditions in the weak field limit\citep{qed-lasers-rev}, where the magnetic field strength is much less than the critical magnetic field strength, $B\ll B_Q$. In contrast, significantly stronger magnetic fields are present in nature. In particular, neutron stars can possess magnetic fields exceeding the Schwinger field and reaching $B_* \sim 10^{15}$ gauss and more. This category of neutron stars is designated as magnetars \citep{magnetars-rev}. At QED field strengths Maxwell's equations become nonlinear, which leads to a wealth of interesting phenomena, such as, polarization-dependent vacuum index of refraction (vacuum birefringence), scattering of light by light, and three-photon interactions (photon splitting), including applications to neutron star magnetospheres \citep{scatt-light-light, photon-split-adler, landavshitz, DGbook, HL06, P+04}. In the recent years, the QED effects were included in computational codes at the level of single-particle effects and are used, for instance, in modeling magnetic reconnection in ultra-strongly magnetized plasmas and the generation of QED cascades in laser plasmas \cite{qed-solver,qed-cascades-laser-17,qed-reconn-19,qed-reconn-23}, as well as in a general QED-plasma solver \citep{Benacek+25}.

Notably, however, the electromagnetic fields in magnetars and laser experiments are not vacuum fields. In contrast, the systems possess a substantial plasma component. In laser experiments, plasma is either created by the interaction with a target, or the $e^{\pm}$ pair plasma can be created from in the laser beam via the Breit-Wheeler process \cite{positron-from-laser-exp,breight-wheeler-plasma-from-light}.
In magnetars, the magnetosphere can twisted by surface shear motions\citep{magnetars-rev,BT07}, so it is threaded by electric currents ${\bm j}=(c/4\pi){\bm \nabla}\times{\bm B}$. Thus, their magnetospheres carry electron-positron plasma needed to maintain the current, $n=j/c e\simeq 10^{17} B_{15}r_{6}^{-1}{\rm cm}^{-3}$, where $B_{15}=B/(10^{15}\textrm{ gauss})$ and $r_{6}=r/(10^{6}\textrm{ cm})$ is the radial distance from the center of the magnetar. Quite importantly, the magnetospheric plasma is not completely charge-neutral. In order to maintain corotation with the star (the magnetic field is anchored in the neutron star crust), the magnetosphere must locally contain plasma with a charge density at least equal to the ``Goldreich-Julian" charge density, $\rho_{\rm GJ}\approx{{\bm\Omega}_*\cdot {\bm B}}/{2c}$, where $\bm\Omega_*$ is the angular velocity of the neutron star (magnetar) rotation. Note that the sign of the charge density varies throughout the magnetosphere depending on the angle between the magnetic field and the neutron star spin. The ``Goldreich-Julian'' particle density is $n_{\rm GJ}=\rho_{\rm GJ}/e\simeq(7\times10^{13}\textrm{ cm}^{-3}){B_{15}}/{P}$, where $P=2\pi/\Omega_*$ is the spin period of the magnetar or a neutron star in seconds. The ratio ${\cal M}=n/n_{\rm GJ}\sim10^3r_6^{-1}$ is called the ``multiplicity'' and is related to the degree of non-neutrality of the magnetospheric plasma (see below). Finally, a neutron star (magnetar) crust is a conducting solid with free electrons, thus, it is also a magnetized purely electron plasma. The electron density varies between $n\sim 2.5\times10^{36}\textrm{~cm}^{-3}$ at the liquid core--solid crust interface and $n\sim 3\times 10^{31}\textrm{~cm}^{-3}$ at the neutron star surface \citep{pearson+11,potekhin+13}. 

Paper I \citep{M23} develops the QED-plasma framework, which incorporates the nonlinear QED-Maxwell equations into plasma dynamics. It also considers how the normal modes in a cold plasma ($k_B T\to 0$) are modified by QED effects. 
\newtext{
This paper, hence referred to as Paper II, extends the analysis from Paper I
to the the case of relativistic plasma,
which temperature is much larger than the particles' rest mass $k_B T\gg m_ec^2$.
Therefore, we briefly summarize the main results from Paper I here.
}

\newtext{
The electromagnetic field equations are obtained from demanding that
the action is stationary under variation, that is
$\delta S = \delta \int \mathcal{L}\ dV \ dt$.
The classical Lagrangian density for the fields and matter-field interaction
yields Maxwell's equations.
However, under QED the Lagrangian contains additional corrections
so that $\mathcal{L} = \mathcal{L}_{\text{classical}} + \mathcal{L}_{\text{QED}}$.
The one-loop approximation is given by \citep{scatt-light-light,Heisenberg1936,Weisskopf1936}
\begin{align}
    \mathcal{L}_{\text{QED,1-loop}} = \frac{m_e c^2}{8\pi^2}\left(\frac{m_e c}{\hbar}\right)^3 \int_0^\infty \frac{e^{-\eta}}{\eta^3} \nonumber\\
    \times\left[
        -\left(\eta a \cot \eta a\right) \left(\eta b \coth \eta b\right)
        + 1 - \frac{\eta^2}{3}\left(a^2-b^2\right)
    \right] d\eta,
\end{align}
where the parameters $a$ and $b$ are
\begin{equation}
    a = \frac{\hbar e E}{m_e^2 c^3} \equiv \frac{E}{E_Q},
\end{equation}
\begin{equation}
    b = \frac{\hbar e B}{m_e^2 c^3} \equiv \frac{B}{B_Q},
\end{equation}
respectively.
The nonlinear Maxwell's equations result upon
variation with respect to the fields.
In an arbitrarily strong magnetic field, the modified field equations
yield corrections through the electric and magnetic
vacuum susceptibilities, $\chi_{ij}^{\rm vac}$ and $\eta_{ij}^{\rm vac}$.
In particular, corrections arise from the quantities
$C_\delta$, $C_\epsilon$, and $C_\mu$,
which are complicated functions of $B/B_Q$
and are discussed in Section \ref{s:linear}.
The QED-plasma framework is obtained by linearizing
the QED corrected field equations to analyze
plasma normal modes.
}

\newtext{
In Paper I, the QED-plasma framework was applied to
a cold electron-positrion plasma,
where the plasma temperature is small compared
to the first excited Landau level,
$kT_\parallel \ll \hbar \Omega$,
and $\Omega = eB/m_e c$ is the electron cyclotron frequency.
The dispersion relation was obtained, which allowed
for the analysis of resonances, cutoffs, and the
general mode structure.
It was found that the general mode structure remains the same
as in a classical plasma. QED effects did not introduce any new modes.
QED has a global effect through renormalization of the plasma frequency
\begin{equation}
    \omega_{p*} = \frac{\omega_p}{1-C_\delta},
\end{equation}
where $\omega_{p}=\sqrt{4\pi n e^2/m_e}$ the non-relativistic electron plasma frequency.
In addition to this renormalization, QED effects only
enter into the dispersion relation via
two functions, $\alpha_\epsilon$ and $\alpha_\mu$,
which are defined in Eq. \ref{alphaem} and
displayed as a function of $B/B_Q$ in Fig. \ref{f:waa}.
Due to these effects, cutoff and resonance frequencies,
$\omega_0$ and $\omega_\infty$ respectively,
become dependent on $B$-field and propagation angle $\theta$.
We highlight the most prominent of these changes.
The O-mode cutoff frequency is reduced,
\begin{equation}
    \omega_{0}^{\left(1\right)}=\frac{\omega_{p*}}{\sqrt{1+\alpha_{\epsilon}}},
\label{eq:OmodeCold}
\end{equation}
allowing the O-mode to propagate at frequencies below
the plasma frequency, where $\alpha_\epsilon={C_{\epsilon}}/{1-C_{\delta}}$.
The Alfv\'en mode resonance is reduced at high $\omega$, $k$ by
\begin{equation}
    \omega_\infty^{\left(2\right)} \approx 
    \frac{\omega_{p*}\cos\theta}{\sqrt{1+\alpha_\epsilon \cos^2 \theta}}.
\end{equation}
The ordinary mode is slowed through the increase in
its index of refraction.
For $\omega\gg\omega_p$, the dispersion relation is
\begin{equation}
    \omega=kc\sqrt{\frac{1+\alpha_\epsilon \cos^2\theta}{1+\alpha_\epsilon}}.
\end{equation}
These general effects can be seen in the cold plasma dispersion curves
in Fig. \ref{fig:dispersion-compare-QED}.
}

\newtext{We are now interested in the case where the plasma temperature
is non-negligible, in particular the case where temperature is much larger than the particles' rest mass $k_B T\gg m_ec^2$. Thus, we will now extend the results from Paper I to a plasma with these thermal effects.}
The rest of the paper is organized as follows. Section \ref{s:linear} describes the linear wave formalism in a relativistic ultra-magnetized plasma and derives the corresponding dispersion relation. In Section \ref{s:analysis} a comprehensive analytical analysis of the normal mode dispersion, their characteristic frequencies and relations is presented. Section \ref{s:num} presents the full numerical solution of the dispersion relation and identifies the corresponding branches. Their dependence on the field strength and temperature is illustrated. Section \ref{s:summary} summarizes essential results.

\section{Linear waves} 
\label{s:linear}
 
In this study, we employ the QED-plasma framework developed in Paper I \citep{M23}. This framework enables the investigation of the normal plasma modes, provided with the plasma electric susceptibility tensor, $\chi_{ij}^{\rm plasma}=\epsilon_{ij}-\delta_{ij}$, where $\epsilon_{ij}$ is the plasma dielectric tensor. The reader is referred to Paper I for extensive details. 

Consider a non-neutral ultrarelativistic pair plasma characterized by the non-neutrality parameter $\Delta n/n$, where $\Delta n=n^+-n^-$ is the difference between the background positron and electron plasma densities, and $n_0=n^++n^-$ is the total lepton density. We, however, assume that there are no net background currents associated with either species. We consider the case of strongly magnetized and magnetically dominated plasma, so that the cyclotron frequency is much larger than the electron plasma frequency, and the electron plasma frequency is much larger than the frequency of the considered wave modes. 
%The relativistic versions of the cyclotron and plasma frequencies depend on the details of the particle distribution functions; they will be specified below.  
%They are related to their non-relativistic counterparts approximately as $\Omega_{e, rel}\sim \Omega_e/\langle\gamma\rangle$ and $\omega^2_{pe, rel}\sim \omega^2_{pe}\langle 2/\gamma^3\rangle$, 

In what follows, we denote $\omega_{p}=\sqrt{4\pi n e^2/m_e}$ the non-relativistic electron plasma frequency and $\Omega=eB/m_ec$ the non-relativistic electron cyclotron frequency. The electron gyroscale  in a QED-strong magnetic field is negligibly small,  $k_\perp^2\rho_{e}^2\ll 1$, because the electrons reach the lowest energy Landau level almost instantaneously.  
We conventionally choose the coordinate frame such that the wave vector is given by ${\bf k}=(k_{\perp},0,k_{z})$, where $z$ is the direction of the background magnetic field. 
\newtext{For oblique propagation, waves propagating at angle $\theta$ with respect to the field, this
gives components $\mathbf{k} = \left(k \sin\theta,0, k\cos\theta\right)$.}

Under these assumptions, the plasma electric susceptibility tensor is given by expression \cite[e.g.,][]{M23,gedalin1998,gedalin2001}:
\begin{equation}
\chi_{ij}^{\rm plasma}=\begin{pmatrix}
\chi_\bot & i\,g & 0 \\
-i\,g & \chi_\bot & 0 \\
0 & 0 & \chi_\| \\
\end{pmatrix},
\end{equation} 
where
\begin{align}
\chi_\bot &= -\frac{\omega_{p}^2}{\omega^2-\Omega^2} ,\\
\chi_\| &= -Q(\omega, {\bm k}), \\
g&= -\frac{\omega_{p}^2\,\Omega}{\omega\left(\omega^2-\Omega^2\right)}\frac{\Delta n}{n}.
\label{g*}
\end{align}
Note that the off-diagonal components $\pm ig$ are related to the breakdown of the charge neutrality. If we assume that $n_{\rm GJ}$ represent the entirely non-neutral fraction of the plasma, then the ``non-neutrality fraction'', $\Delta n/n$, that will appear later can be approximated as the inverse multiplicity factor $\Delta n/n\sim n_{\rm GJ}/n\sim {\cal M}^{-1}$. 

In the above expressions, the function $Q(\omega, {\bm k})$ depends on the particle distribution function. It will be discussed below. Note the shape of the particle distribution and the thermal effects in particular enter the plasma dispersion in the direction parallel to the magnetic field only. In the perpendicular plane, the plasma is cold and resides in the lowest Landau level. 

The QED effects modify the Maxwell equations and make them nonlinear. They introduce additional electric and magnetic susceptibilities of vacuum:
\begin{align}
\chi_{ij}^{\rm vac}&=-(C_{\delta}\delta_{ij}-C_{\epsilon}b_{i}b_{j}),\label{chi}\\
\eta_{ij}^{\rm vac}&=-(C_{\delta}\delta_{ij}+C_{\mu}b_{i}b_{j}).\label{eta}
\end{align}
Here the coefficients $C_\delta$, $C_\epsilon$, and $C_\mu$ are given by complicated functions of $B/B_{Q}$ presented by Eqs. (28--30) and shown in Fig. 1 in Paper I \citep{M23}. 

\newtext{The physical meaning of these $C$-coefficients is that they represent the modification of the vacuum permittivity and inverse permeability, which  become non-unity and anisotropic:
\begin{align}
\epsilon_{ij}^{\rm vac}&=\delta_{ij}-(C_{\delta}\delta_{ij}-C_{\epsilon}b_{i}b_{j}),\label{eps}\\
\mu_{ij}^{-1,\rm vac}&=\delta_{ij}-(C_{\delta}\delta_{ij}+C_{\mu}b_{i}b_{j}).\label{mu}
\end{align}
Consequently, $C_{\delta}$ represents the modification of the isotropic part of the susceptibilities, whereas $C_{\epsilon}$ and $C_{\mu}$ represent anisotropic contributions to the electric and magnetic susceptibilities. }

Since the expressions are long and cumbersome, they are not presented here. 
However, the approximate scalings for both $B\ll B_Q$ and $B\gg B_Q$ are rather simple\citep{M23}. In the weak field limit, $B\ll B_{Q}$, these quantities take the  values
\begin{align}
C_{\delta} & =(2/45)\alpha (B/B_{Q})^{2},\label{cdweak}\\
C_{\epsilon} & =(4/45)\alpha (B/B_{Q})^{2},\label{ceweak}\\
C_{\mu} & =(7/45)\alpha (B/B_{Q})^{2}.\label{cmweak}
\end{align}
In contrast, in the very strong field limit, $B\gg B_{Q}$, the quantities scale as
\begin{align}
C_{\delta} & \propto \log(B/B_{Q}),\label{cdstrong}\\
C_{\epsilon} & \propto (B/B_{Q}),\label{cestrong}\\
C_{\mu} & \sim const.\label{cmstrong}
\end{align}

The total electric permittivity and inverse magnetic permeability are given by equations:
\begin{align}
\epsilon_{ij}&=\delta_{ij}+\chi_{ij}^{\rm vac}+\chi_{ij}^{\rm plasma},\\
\mu_{ij}^{-1}&=\delta_{ij}+\eta_{ij}^{\rm vac}.
\end{align}

The dispersion relations and polarizations of the plasma waves are found from the wave equation:
\begin{equation}
\left(\frac{\omega^2}{c^2}\epsilon_{il}-e_{ijk}e_{lrq} k_j k_r\mu_{kq}^{-1}\right)\tilde E_l=0
\end{equation}
or more explicitly 
\begin{align}
&\left[\frac{\omega^2}{c^2}\epsilon_{il}+\mu^{-1}_{il}k^{2}-\mu^{-1}\left(\delta_{il}k^2-k_i k_l\right)
\right. \nonumber\\
& \left.\quad\phantom{\frac{\omega^2}{c^2}}
+\delta_{il}\mu^{-1}_{jk}k_j k_k-\mu^{-1}_{ij}k_j k_l -\mu^{-1}_{lj}k_j k_i\right]\tilde E_l=0,
\end{align}
where $\tilde E_l$ denotes the fluctuating electric field of a wave, $e_{ijk}$ is the Levi-Civita symbol, $\mu^{-1}\equiv \mu^{-1}_{ii}$ is the trace of $\mu^{-1}_{ij}$, and $k^2=k_i k_i$. 

Equating the determinant of the matrix in the square brackets to zero, we obtain:
\begin{align}
&\textrm{det}\left[
\frac{\omega^2}{c^2}\left(\delta_{ij}+\chi_{ij}^{\rm vac}+ \chi_{ij}^{\rm plasma}\right)-\left(\delta_{ij}k^2-k_i k_j\right)\left(1+\eta^{\rm vac}\right)
\right.\nonumber\\
& \left.\phantom{\frac{\omega^2}{c^2}}
+\delta_{ij}\eta_{mn}^{\rm vac}k_m k_n+\eta^{\rm vac}_{ij}k^2-\left(\eta_{im}^{\rm vac} k_m k_j+\eta_{jm}^{\rm vac}k_m k_i\right)
\right]=0,
\label{disp-long}
\end{align}
where $\eta^{\rm vac}=\eta_{ii}^{\rm vac}$ is the trace of $\eta_{ij}^{\rm vac}$.
\newtext{Recalling our choice of frame where
$\mathbf{k} = \left(k \sin\theta,0, k\cos\theta\right)$,}
upon some algebra, the dispersion equation takes the following form:
%\begin{strip}
\begin{widetext}
\begin{align}
&{\rm det}\left[\frac{\omega^2}{c^2}
\begin{pmatrix}
1-C_{\delta} & 0 & 0 \\
0 & 1-C_{\delta} & 0 \\
0 & 0 & 1-C_{\delta}+C_{\epsilon}
\end{pmatrix}\right.
+\frac{\omega^2}{c^2}\Bigg(\chi_{ij}^{\rm plasma}\Bigg)
\nonumber\\
&\qquad\left.-k^2\begin{pmatrix}
\cos^2\theta\left(1-C_{\delta}\right) & 0 & \sin\theta\cos\theta \left(1-C_{\delta}\right) \\
0 & \left(1-C_{\delta}-C_{\mu}\sin^2\theta\right) & 0 \\
\sin\theta\cos\theta\left(1-C_{\delta}\right) & 0 & \sin^2\theta\left(1-C_{\delta}\right)
\end{pmatrix}\right] =0.
\label{disp-main}
\end{align}
\end{widetext}

In order to proceed further, we need to specify the function~$Q$. If we consider a one-dimensional particle velocity distribution function, we get \cite[e.g.,][]{gedalin1998,vega2024}:
\begin{eqnarray}
\label{Q}
Q(\omega, {\bm k})=\frac{\omega_{p}^2}{\omega^2}\,W\left(\omega, k_z\right),
\end{eqnarray}
where the $W$ function is given by the standard integral:
\begin{eqnarray}
W=-\frac{\omega^2}{k_z}\int\limits_{-c}^{c}\frac{1}{\omega-k_zv_z +i\nu}\,\frac{df}{dv_z}\,dv_z.
\label{W}
\end{eqnarray}
Here, $\nu\to +0$ is needed to take into account collisionless Landau damping. The distribution function is typically expressed through the variable $u_z=v_z/\sqrt{1-v_z^2/c^2}$, so that its normalization takes a simple form:
\begin{eqnarray}
\int\limits_{-\infty}^{\infty}f(u_z)du_z=1.    
\end{eqnarray}
For simplicity, we consider the one-dimentional Maxwell-J{\"u}ttner distribution,
\begin{eqnarray}
f(u_z)=\frac{1}{2cK_1(1/\Theta)}\exp{\left(-\gamma/\Theta\right)},    
\end{eqnarray}
where $K_1$ is the modified Bessel function of the second kind, $\gamma=1/\sqrt{1-v_z^2/c^2}=\sqrt{1+u_z^2/c^2}$ is the relativistic gamma-factor, and $\Theta=k_BT/(m_ec^2)$ is the temperature parameter. We will consider the ultrarelativistic limit, $\Theta\gg 1$, in which case $K_1(1/\Theta)\approx \Theta$.

The Landau damping is strong when $1-\omega^2/(k_z^2c^2)\sim 1/\Theta^2$, which is the condition when the phase velocity of the wave is comparable to the thermal velocities of the particles.  In this case, the function $Q$ has a large imaginary part, comparable to its real part. Its imaginary part is relatively small in the following two limiting cases:
\begin{eqnarray}
\label{case1}
\left|1-\frac{\omega^2}{k_z^2 c^2}\right| \gg \frac{1}{\Theta^2}\quad \mbox{(Case I)},    
\end{eqnarray}
and 
\begin{eqnarray}
\label{case2}
\left| 1-\frac{\omega^2}{k_z^2 c^2}\right| \ll \frac{1}{\Theta^2}\quad \mbox{(Case II)}.    
\end{eqnarray}
As discussed in \cite[e.g.,][]{godfrey1975,gedalin1998,vega2024}, in Case I, the $Q$ function takes the form
\begin{eqnarray}
Q\approx \frac{\omega_{p}^2}{\Theta\left(\omega^2-k_z^2c^2\right)},    
\end{eqnarray}
while in Case II, we have
\begin{eqnarray}
Q\approx \frac{2\Theta \omega_{p}^2}{k_z^2c^2}.    
\end{eqnarray}
Obviously, in the case $\omega^2\geq k_z^2 c^2$, the imaginary part is absent since the phase velocity of such waves is greater than the speed of light, and they are not affected by Landau damping.

\begin{figure*}[t]
%\vskip0.5cm
\includegraphics[scale = 0.95]{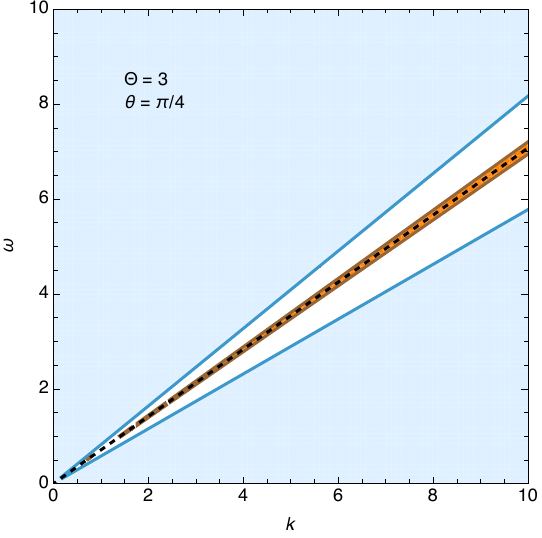}  %% file name without extension
\includegraphics[scale = 0.95]{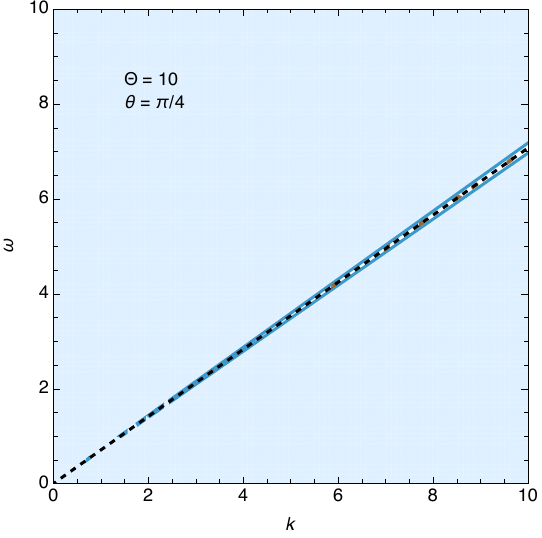}  %% file name without extension
\caption[]{Regions approximately corresponding to Case I (light-blue) and Case II (orange) are shown for two values of temperature $\Theta$ and two propagation angles $\theta$. Above the dashed black line, waves are superluminal and decoupled from Landau damping. The damping can be significant in the unshaded region below the dashed line, in-between the blue and orange regions. Case I  dominates waves' dispersion, especially in the ultra-relativistic plasma. 
}
\label{f:allow}
\end{figure*}

Fig. \ref{f:allow} illustrates the location of regions corresponding to Case I and Case II in the $k$-$\omega$ plane. For concreteness, we assume that a quantity is ``much greater'' or ``much smaller'' when it differs by a factor of three. The big light-blue region represents inequality $\left|1-{\omega^2}/{k_z^2 c^2}\right| > {3}/{\Theta^2}$, and the narrow orange region represents $\left|1-{\omega^2}/{k_z^2 c^2}\right| < {1}/{3\Theta^2}$. The region above the dashed black line is where the waves are superluminal, $\omega^2\geq k_z^2 c^2$, hence Landau damping is absent. The collisionless damping can be important in the unshaded region below the dashed line, in-between the blue and orange regions. It is seen that Case I dominates the waves' dispersion, especially in the ultra-relativistic regime, $\Theta\gg1$.

\begin{figure*}[t]
%\vskip0.5cm
\includegraphics[scale = 1.4]{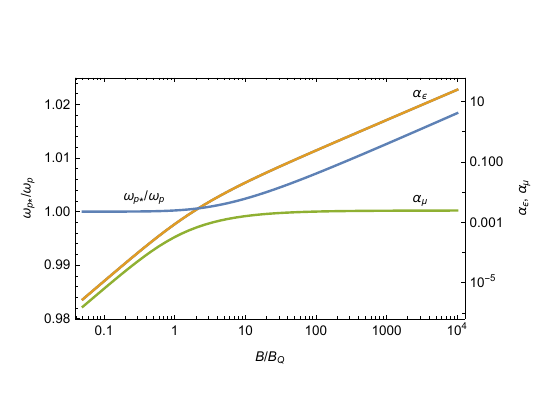}  %% file name without extension
\caption[]{QED modification of the plasma frequency (log-linear scale, left axis) and quantities $\alpha_{\epsilon}, \alpha_{\mu}$ (log-log scale, right axis) as a function of the magnetic field strength $B/B_{Q}$.}
\label{f:waa}
\end{figure*}

As in Paper I \citep{M23}, we observe that (i) Eq. (\ref{disp-long}) contains a common term $(1-C_{\delta})$ and (ii) all plasma susceptibilities $\chi_{ij}$ are proportional to $\omega_{p}^{2}$. Therefore, we renormalize the plasma frequency and define new quantities:
\begin{equation}
\omega_p^2\to\omega_{p*}^2\equiv \frac{\omega_p^2}{1-C_{\delta}},
\quad
\alpha_\epsilon=\frac{C_{\epsilon}}{1-C_{\delta}}, 
\quad
\alpha_\mu=\frac{C_{\mu}}{1-C_{\delta}}.
\label{alphaem}
\end{equation}
Fig. \ref{f:waa} illustrates the behavior of these parameters as a function of the field strength. \newtext{We remind that the physical meaning of the coupling $C$-coefficients is that they represent the strong-field QED-induced modification of the vacuum susceptibilities, as is described in Eqs. \eqref{eps}, \eqref{mu}. Consequently, the physical meaning of $\alpha_\epsilon$ and $\alpha_\mu$ is that these parameters represent the normalized anisotropic modifications of the vacuum permittivity and permeability, respectively. }

With the above definitions, the dispersion relation is
%\begin{strip}
\begin{widetext}
\begin{equation}
{\rm det}\left[
\frac{\omega^2}{c^2}\begin{pmatrix}
\epsilon_{\bot*} & i\,g_* & 0 \\
-i\,g_* & \epsilon_{\bot*} & 0 \\
0 & 0 & \epsilon_{\|*} \\
\end{pmatrix}
-k^2\begin{pmatrix}
\cos^2\theta & 0 & \sin\theta\cos\theta \\
0 & 1-\alpha_\mu\sin^2\theta & 0 \\
\sin\theta\cos\theta & 0 & \sin^2\theta
\end{pmatrix}\right] =0,
\end{equation}
\end{widetext}
%\end{strip}
where
%\begin{displaystyle}
\begin{align}
\epsilon_{\bot*} &= 1-\frac{\omega_{p*}^2}{\omega^2-\Omega^2}, \\
\epsilon_{\|*} &=  1+\alpha_\epsilon-Q_*, \\
g_*&=  -\frac{\omega_{p*}^2\,\Omega}{\omega\left(\omega^2-\Omega^2\right)}\frac{\Delta n}{n}, \\
Q_*&\approx
\left\{
\begin{array}{ll}
\displaystyle
\frac{\omega_{p*}^2}{\Theta\left(\omega^2-k^2c^2\cos^{2}\theta\right)} &  \textrm{(Case I)}, \\
\displaystyle
\frac{2\Theta \omega_{p*}^2}{k^2c^2\cos^{2}\theta}  &  \textrm{(Case II)}.
\end{array}\right.
\end{align}
%\end{displaystyle}

One readily sees that the effect of the quantum vacuum reduces to the renormalization of the plasma frequency and addition of two field-dependent coefficients to the dispersion relation, via $\alpha_\epsilon$ in the $\omega^{2}$-term and via $\alpha_\mu$ in the $\epsilon_{\|*}$ component entering the $k^{2}$-term. In a super-critical magnetic field $B\gg B_{Q}$, the only strong effect is due to $\alpha_\epsilon$, which grows linearly with the field strength $\alpha_\epsilon\propto B$. It exceeds unity $\alpha_\epsilon>1$ in the field $B/B_{Q}\gg 1/\alpha\sim 137$. The QED modification to the plasma frequency is small, on the order of a few percent. It grows logarithmically with the field strength $(\omega_{p*}-\omega_{p})/\omega_{p}\simeq C_{\delta}\propto \log B$. The contribution from $\alpha_\mu$ is always small and saturates $\alpha_\mu\sim \textrm{few}\times10^{-3}$.

Introducing the index of refraction, $N^2=k^2 c^2/\omega^2$, we finally obtain 

%\begin{strip}
\begin{widetext}
\begin{equation}
{\rm det}
\begin{bmatrix}
N^2\cos^2\theta -\epsilon_{\bot*} & -i\,g_* & N^2\sin\theta\cos\theta \\
i\,g_* & N^2\left(1-\alpha_\mu\sin^2\theta\right)-\epsilon_{\bot*} & 0 \\
N^2\sin\theta\cos\theta & 0 & N^2\sin^2\theta-\epsilon_{\|*}
\end{bmatrix} = 0.
\label{eq:disp-matrix}
\end{equation}
\end{widetext}
%\end{strip}
Expansion of the determinant yields
\begin{equation}
N^4 \mathsf{A} + N^2 \mathsf{B} + \mathsf{C}=0,
\label{N-disp}
\end{equation}
where the scalar coefficients, $\mathsf{A},\ \mathsf{B},\ \mathsf{C}$, are 
\begin{align}
\mathsf{A}&= \left(\epsilon_{\bot*}\sin^2\theta+\epsilon_{\|*}\cos^2\theta\right) 
\left(1-\alpha_\mu\sin^2\theta\right), 
\label{A}\\
\mathsf{B}&= -\left[\epsilon_{\bot*}\,\epsilon_{\|*}
\left(1+\cos^2\theta-\alpha_\mu\sin^2\theta\right)
+\left(\epsilon_{\bot*}^2-g_*^2\right)\sin^2\theta
\right], 
\label{B}\\
\mathsf{C}&= \epsilon_{\|*} \left(\epsilon_{\bot*}^2-g_*^2\right).
\label{C}
\end{align}
We use the same letter $\mathsf{B}$ for one of the coefficients as for the field strength, hoping this would not cause any confusion. Note that the index of refraction enters $\epsilon_{\|*}$ through the temperature-dependent function
\begin{align}
Q_*&\approx
\left\{
\begin{array}{ll}
\displaystyle
\frac{\omega_{p*}^2}{\Theta\omega^{2}\left(1-N^2\cos^{2}\theta\right)} &  \textrm{(Case I)}, \\
\displaystyle
\frac{2\Theta \omega_{p*}^2}{\omega^{2}N^2\cos^{2}\theta}  &  \textrm{(Case II)},
\end{array}\right.
\end{align}
so that Eq. (\ref{N-disp}) is no longer bi-quadratic. 
By expanding out $\epsilon_{\parallel*}$,
we rearrange the polynomial into the form
\begin{equation}
\left[N^{4}\tilde{A}+N^{2}\tilde{B}+\tilde{C}\right]+Q_{*}\left[N^{4}A_{*}+N^{2}B_{*}+C_{*}\right]=0
\label{eq:N-disp-Q}
\end{equation}
where "starred" quantities are the terms in the coefficients that
are proportional to $Q_*$
and "tilded" quantities are those that are not.
The tilded coefficients are
\begin{flalign}
 & \tilde{A}=\left(\epsilon_{\perp*}\sin^{2}\theta+\left(1+\alpha_{\epsilon}\right)\cos^{2}\theta\right)\left(1-\alpha_{\mu}\sin^{2}\theta\right),\\
 & \tilde{B}=-\left[\epsilon_{\perp*}\left(1+\alpha_{\epsilon}\right)\left(1+\cos^{2}\theta-\alpha_{\mu}\sin^{2}\theta\right)+\left(\epsilon_{\perp*}^{2}-g_{*}^{2}\right)\sin^{2}\theta\right],\\
 & \tilde{C}=\left(1+\alpha_{\epsilon}\right)\left(\epsilon_{\perp*}^{2}-g_{*}^{2}\right),
\end{flalign}
and the starred coefficients are
\begin{flalign}
 & A_{*}=-\cos^{2}\theta\left(1-\alpha_{\mu}\sin^{2}\theta\right),\\
 & B_{*}=\epsilon_{\perp*}\left(1+\cos^{2}\theta-\alpha_{\mu}\sin^{2}\theta\right),\\
 & C_{*}=-\left(\epsilon_{\perp*}^{2}-g_{*}^{2}\right).
\end{flalign}
As $Q_*$ contains factors of $N^2$ in the denominator,
we can rearrange Eq. (\ref{eq:N-disp-Q})
into a bi-cubic polynomial
\begin{equation}
\mathcal{A}N^{6}+\mathcal{B}N^{4}+\mathcal{C}N^{2}+\mathcal{D}=0
\label{eq:N-cubic}
\end{equation}
where the coefficients are now combinations of
tilded and starred quantities depending on the
case chosen for $Q_*$.
The coefficients for Case I are listed in Eqs. 
(\ref{eq:casei-coeffs-begin}) - (\ref{eq:casei-coeffs-end}),
while the coefficients for Case II are listed in Eqs.
(\ref{eq:caseii-begin}) - (\ref{eq:caseii-end}).
Thus, the dispersion relation can always be solved in principle
using the cubic formula.
These solutions are typically algebraically complicated.
We now explore simple special cases and
limiting behavior.

\section{Analysis}
\label{s:analysis}

\subsection{Case I}
\begin{figure*}[t]
\includegraphics[scale = 0.9]{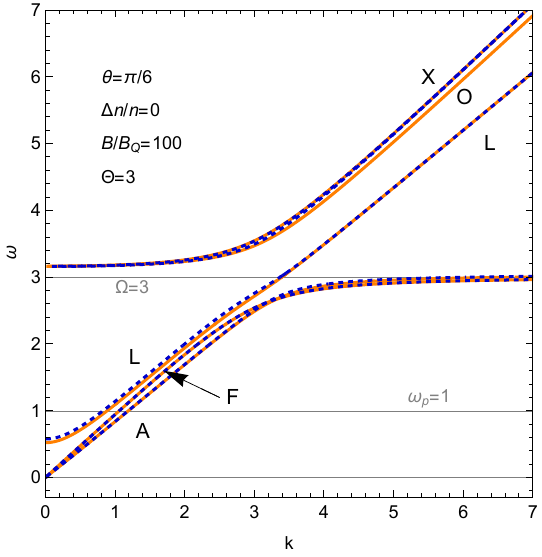}
\includegraphics[scale = 0.9]{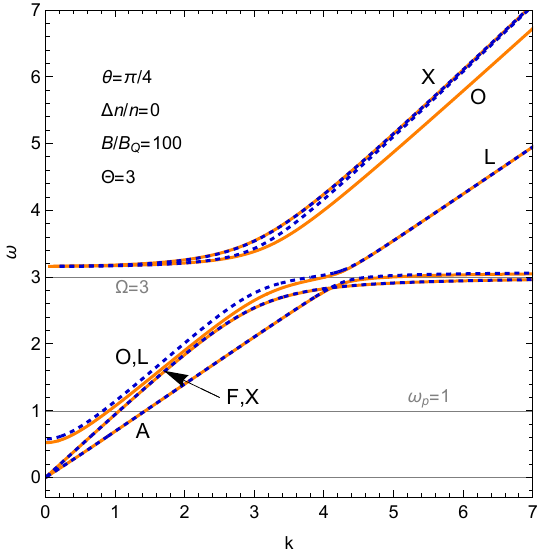}
\includegraphics[scale = 0.9]{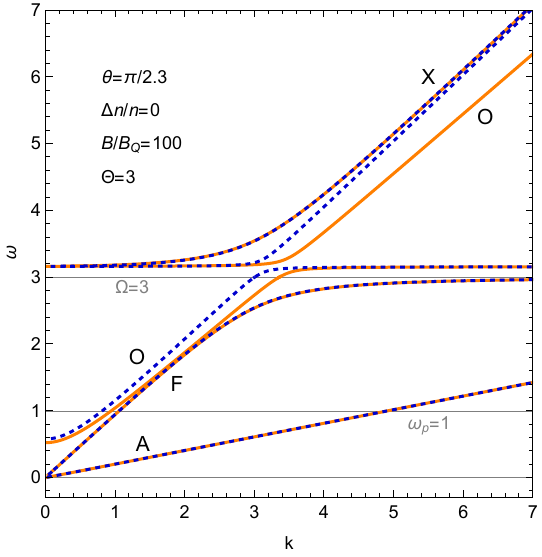}
\includegraphics[scale = 0.9]{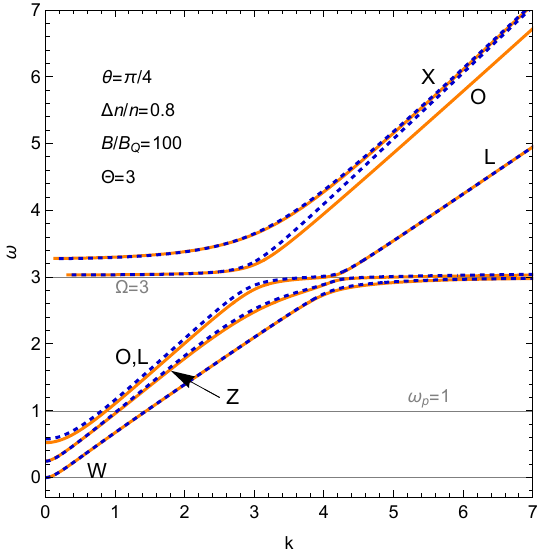}
\caption{The schematic representation the plasma dispersion curves 
$\omega\left(k\right)$ for electrically neutral ($\Delta n/n=0$) and non-neutral ($\Delta n/n=0.8$) relativistic magnetized plasma. 
The units are arbitrary, but we set the speed of light to $c=1$. 
We set the numerical values of the plasma and cyclotron frequencies to be$\omega_p=1$, $\Omega=3$. The temperature parameter is chosen to be $\Theta=3$ for illustrative purposes only as, formally, $\Theta\gg1$ in the ultrarelativistic plasma. Both standard (dashed blue curves) and QED-modified with $B/B_Q=100$ (orange curves) branches of plasma normal modes are shown. 
The wave branches are labeled as follows: 
“A”—Alfv\'en wave, 
“F”—fast magnetosonic wave, 
“X”—extraordinary electromagnetic wave, 
“O”—ordinary electromagnetic wave(in a neutral plasma, it consists of two branches split around the cyclotron frequency),
“W”—whistler wave,
“Z”—Z-mode 
(the lower-frequency branch of the extraordinary wave, also called the slow extraordinary mode),
"L"-Langmuir mode.}
\label{fig:branches}
\end{figure*}

\begin{figure*}[t]

\noindent\begin{minipage}[t]{1\columnwidth}%
\includegraphics[]{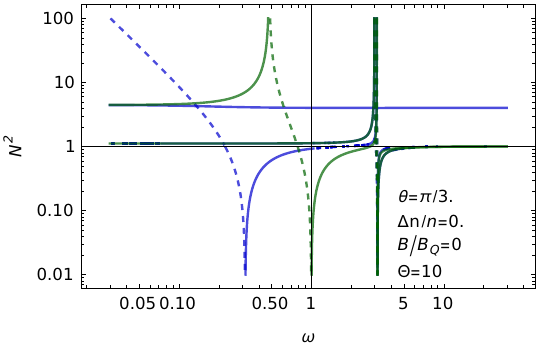}%
\end{minipage}\hfill{}%
\noindent\begin{minipage}[t]{1\columnwidth}%
\includegraphics[]{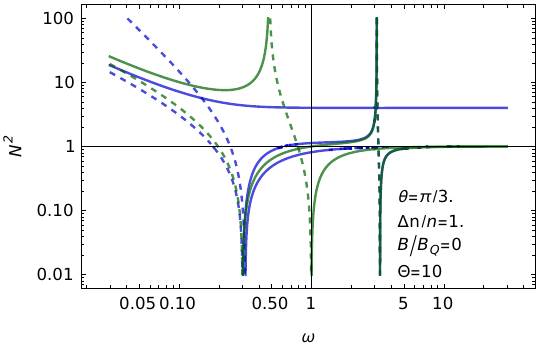}%
\end{minipage}

\noindent\begin{minipage}[t]{1\columnwidth}%
\includegraphics[]{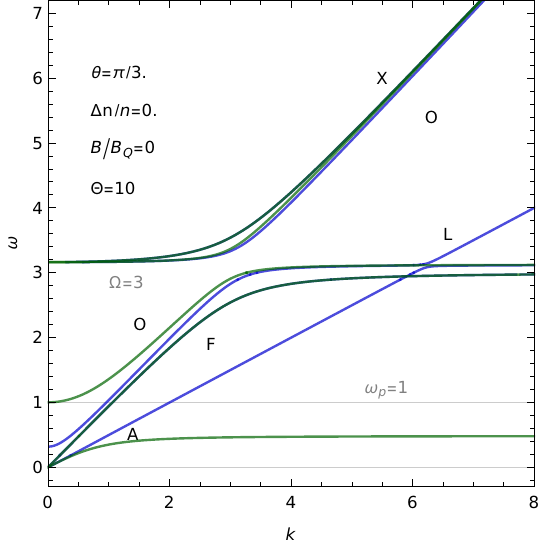}%
\end{minipage}\hfill{}%
\noindent\begin{minipage}[t]{1\columnwidth}%
\includegraphics[]{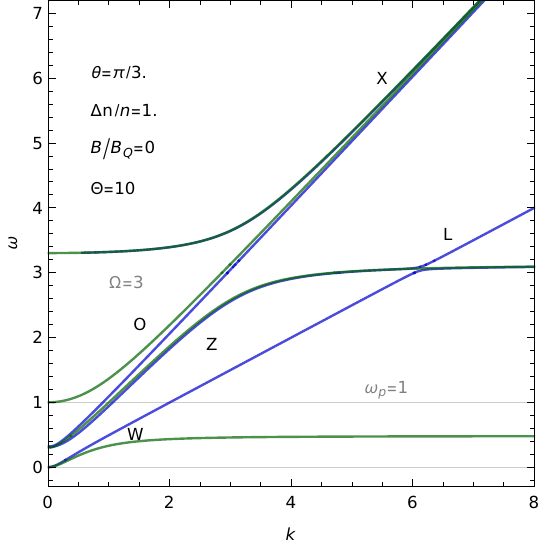}%
\end{minipage}

\caption[]{The schematic representation of the index of refraction squared
$N^2\left(\omega\right)$ (top row) and the plasma dispersion curves 
$\omega\left(k\right)$ (bottom row) 
for electrically neutral and
non-neutral classical plasmas. 
The units are arbitrary, but we set the speed of light to $c=1$. 
We set the numerical values of the plasma and cyclotron frequencies to be
$\omega_p=1$, $\Omega=3$,
and $\theta=\pi/3$.
The cold plasma case is in green, while the
thermal plasma case with $\Theta=10$ is in blue.
Both plots are for the non-QED case with $B/B_Q\rightarrow0$.
Solid lines depict propagating waves, i.e., with $N^2>0$, 
and dashed lines depict evanescent
branches with $N^2<0$.
The wave branches are labeled as follows: 
“A”—Alfv\'en wave, 
“F”—fast magnetosonic wave, 
“X”—extraordinary electromagnetic wave, 
“O”—ordinary electromagnetic wave 
(in a neutral plasma, it consists of two branches split around the cyclotron frequency), 
“W”—whistler wave,
“Z”—Z-mode 
(the lower-frequency branch of the extraordinary wave, also called the slow extraordinary mode),
"L"-Langmuir mode.}
\label{fig:dispersion-compare-classical}
\end{figure*}

\begin{figure*}[t]

\noindent\begin{minipage}[t]{1\columnwidth}%
\includegraphics[]{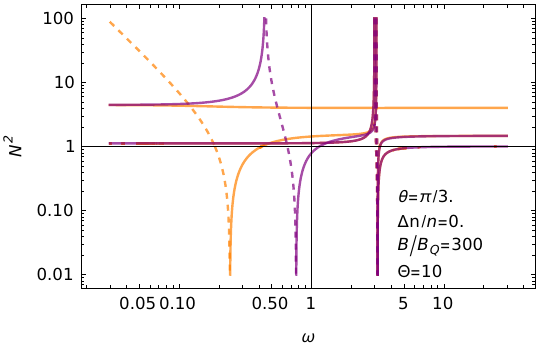}%
\end{minipage}\hfill{}%
\noindent\begin{minipage}[t]{1\columnwidth}%
\includegraphics[]{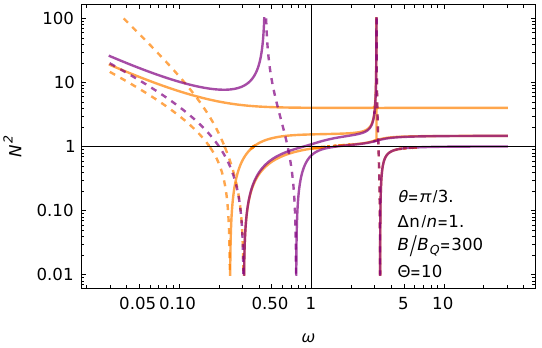}%
\end{minipage}

\noindent\begin{minipage}[t]{1\columnwidth}%
\includegraphics[]{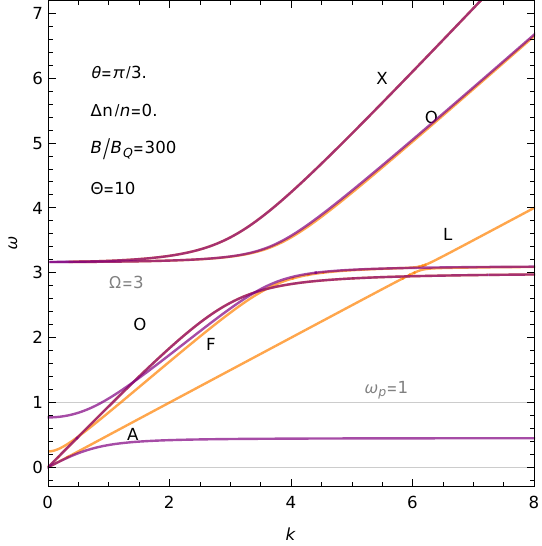}%
\end{minipage}\hfill{}%
\noindent\begin{minipage}[t]{1\columnwidth}%
\includegraphics[]{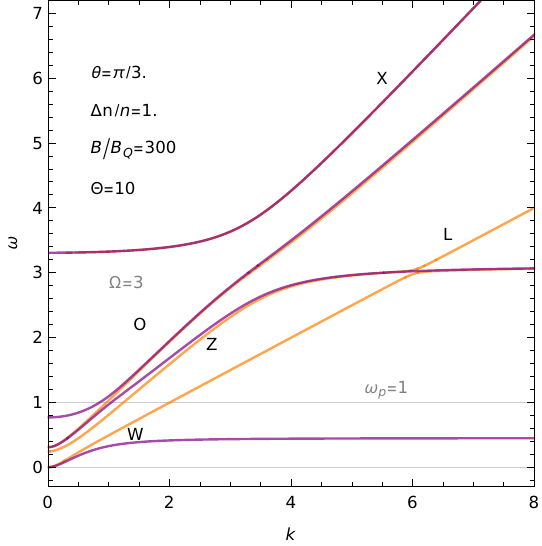}%
\end{minipage}

\caption[]{The schematic representation of the index of refraction squared
$N^2\left(\omega\right)$ (top row) and the plasma dispersion curves 
$\omega\left(k\right)$ (bottom row) 
for electrically neutral and
non-neutral QED plasmas. 
The units are arbitrary, but we set the speed of light to $c=1$. 
We set the numerical values of the plasma and cyclotron frequencies to be
$\omega_p=1$, $\Omega=3$,
and $\theta=\pi/3$.
The wave branches are labeled as in Fig. \ref{fig:dispersion-compare-classical}.
The cold plasma case is in purple, while the
thermal plasma case with $\Theta=10$ is in orange.
Both plots are for the QED case with $B/B_Q=300$.
Solid lines depict propagating waves, i.e., with $N^2>0$, 
and dashed lines depict evanescent
branches with $N^2<0$.
}
\label{fig:dispersion-compare-QED}
\end{figure*}

In Case I, the coefficients in Eq (\ref{eq:N-cubic}) are
\begin{flalign}
 & \mathcal{A}=-\cos^{2}\theta\frac{\Theta\omega^{2}}{\omega_{p*}^{2}}\tilde{A} \label{eq:casei-coeffs-begin}, \\
 & \mathcal{B}=\frac{\Theta\omega^{2}}{\omega_{p*}^{2}}\left(\tilde{A}-\cos^{2}\theta\tilde{B}\right)+A_{*},\\
 & \mathcal{C}=\frac{\Theta\omega^{2}}{\omega_{p*}^{2}}\left(\tilde{B}-\cos^{2}\theta\tilde{C}\right)+B_{*},\\
 & \mathcal{D}=\frac{\Theta\omega^{2}}{\omega_{p*}^{2}}\tilde{C}+C_{*}.
 \label{eq:casei-coeffs-end}
\end{flalign}
The majority of parameter space is occupied by Case I (Fig. \ref{f:allow}), especially
in the ultra-relativistic regime. In particular, the asymptotic behavior of
the dispersion relation will mainly reside in Case I with the only exceptions
being when $\omega \sim k_z$. 
We present the general behavior in Figs. \ref{fig:branches}
- \ref{fig:dispersion-compare-QED}.
Analytic results can be derived for several limiting cases,
which we explore now.

%%%%%%%%%%%----Classification figures---------
%%%%%%%%%%%%%%%%%%

\subsubsection{Resonances}
When $N^2$ diverges, there are frequencies for which $\omega$ remains constant
as $k \rightarrow \infty$, the resonances. In Case I, $N^2\rightarrow \infty $ when
$\mathcal{A} \rightarrow 0$. This reads
\begin{equation}
-\cos^{2}\theta\frac{\Theta\omega^{2}}{\omega_{p*}^{2}}\tilde{A}=0
\end{equation}
which reduces to
\begin{equation}
\omega_{\infty}^{\left(1\right)}=\sqrt{\Omega^{2}+\frac{\sin^{2}\theta}{\sin^{2}\theta+\left(1+\alpha_{\epsilon}\right)\cos^{2}\theta}\omega_{p*}^{2}}.
\end{equation}
In parallel propagation, $\theta=0$,
this resonance becomes
\begin{equation}
    \omega_{\infty}^{\left(1\right)}=\Omega.
\end{equation}
This branch does \textit{not} appear in this form for perpendicular propagation,
for $\theta=\pi/2$. That is a special case that is handled in 
Section \ref{subsubsec:perp-prop}.

Additional branches still exist when $\mathcal{A}=0$,
which can provide additional resonances.
Consider $\mathcal{B}=0$, which is
\begin{equation}
\frac{\Theta\omega^{2}}{\omega_{p*}^{2}}\left(\tilde{A}-\cos^{2}\theta\tilde{B}\right)+A_{*}=0.
\end{equation}
$\mathcal{A}=0$ implies $\tilde{A}=0$, and after some algebra this reduces to
\begin{multline}
\epsilon_{\perp*}\left(1+\alpha_{\epsilon}\right)\left(1+\cos^{2}\theta-\alpha_{\mu}\sin^{2}\theta\right)\\
-\frac{\omega_{p*}^{2}}{\Theta\omega^{2}}\left(1-\alpha_{\mu}\sin^{2}\theta\right)+\left(\epsilon_{\perp*}^{2}-g_{*}^{2}\right)\sin^{2}\theta=0.
\label{eq:calBzero}
\end{multline}
There is also the possibility of $\mathcal{C}=0$
simultaneously with $\mathcal{B}$,
which adds the condition
\begin{multline}
\epsilon_{\perp*}\left(1+\alpha_{\epsilon}-\frac{\omega_{p*}^{2}}{\Theta\omega^{2}}\right)\left(1+\cos^{2}\theta-\alpha_{\mu}\sin^{2}\theta\right)\\
+\left(\epsilon_{\perp*}^{2}-g_{*}^{2}\right)\left(\sin^{2}\theta+\left(1+\alpha_{\epsilon}\right)\cos^{2}\theta\right)=0.
\label{eq:calCzero}
\end{multline}
Both Eqs. (\ref{eq:calBzero}) and (\ref{eq:calCzero})
share similarities with the analogous condition
for the
cold plasma \cite{M23},
but the extra factor of $N^2$ in
$Q_*$ shuffles around terms between the
coefficients.
The exact behavior of the remaining resonances is
analytically complicated,
but are plotted and explored numerically in 
Figs. (\ref{fig:dispersion-compare-classical}) and (\ref{fig:dispersion-compare-QED}).

\subsubsection{Cutoffs}
Where $N^2 < 0$, the index of refraction is imaginary,
causing waves to rapidly attenuate. The cutoff frequencies are identified
at the boundary $N^2 = 0$. For Case I, this condition is
\begin{equation}
    \tilde{C}+\frac{\omega_{p*}^{2}}{\Theta\omega^{2}}C_{*}=0
\end{equation}
which reads
\begin{equation}
    \left(1+\alpha_{\epsilon}-\frac{\omega_{p*}^{2}}{\Theta\omega^{2}}\right)\left(\epsilon_{\perp*}^{2}-g_{*}^{2}\right)=0.
\end{equation}
This reduces to
\begin{equation}
    \omega_{0}^{\left(1\right)}=\frac{\omega_{p*}}{\sqrt{\Theta\left(1+\alpha_{\epsilon}\right)}}
    \label{eq:cutoff-generalize}
\end{equation}
and
\begin{equation}
    \epsilon_{\perp*}^{2}-g_{*}^{2}=0.
    \label{eq:cutoff-exact}
\end{equation}
Eq. (\ref{eq:cutoff-generalize}) is a generalization
of the cold plasma result,
and Eq. (\ref{eq:cutoff-exact}) is
exactly the cold plasma result \cite{M23}.
Eq. (\ref{eq:cutoff-exact}) has simple limiting cases
for $\left|\Delta n/n\right|\rightarrow0$ and $\left|\Delta n/n\right|\rightarrow1$.
For $\left|\Delta n/n\right|\rightarrow0$, the cutoff frequencies are
\begin{equation}
    \omega_0^{\left(2\right)}\approx \frac{\omega_{p*}^2\Omega}{\omega_{p*}^2+\Omega^2}
    \left|\frac{\Delta n}{n}\right|,
\end{equation}
\begin{equation}
    \omega_0^{\left(3\right)}\approx \sqrt{\omega_{p*}^2+\Omega^2}.
\end{equation}
For $\left|\Delta n/n\right|\rightarrow1$, they are
\begin{equation}
    \omega_0^{\left(2,3\right)}=\sqrt{\omega_{p*}^2+\frac{1}{4}\Omega^2}\mp \frac{1}{2}\Omega.
\end{equation}

\subsubsection{Parallel Propagation}
\label{subsubsec:parallel-prop}
Setting $\theta=0$ in Eq. (\ref{eq:disp-matrix}) leads directly to
three branches. The first corresponds to $\epsilon_{\parallel*}=0$, which for
Case I works out to be
\begin{equation}
    N^2 = 1 - \frac{\omega_{p*}^2}{\Theta \omega^2 \left(1+\alpha_\epsilon\right)}.
\end{equation}
This is the Langmuir mode branch with dispersion relation
\begin{equation}
    \omega^2 = k^2c^2 + \frac{\omega_{p*}^2}{\Theta \left(1+\alpha_\epsilon\right)}.
\end{equation}
The remaining branches are shared
between both cases, and are obtained from
\begin{equation}
    \left(N^2 - \epsilon_{\perp*}\right)^2 - g_*^2=0,
\end{equation}
which has solutions
\begin{equation}
    N^2 = \epsilon_{\perp*}\pm g_*.
    \label{eq:parallel-shared}
\end{equation}
This case is similar to the cold plasma of Paper I \citep{M23}.

\subsubsection{Perpendicular Propagation}
\label{subsubsec:perp-prop}
At $\theta=\pi/2$, Eq. (\ref{eq:N-cubic}) reduces to
the bi-quadratic equation
\begin{equation}
    N^4\left(\tilde{A}+\frac{\omega_{p*}^2}{\Theta\omega^2}A_*\right) +
    N^2\left(\tilde{B}+\frac{\omega_{p*}^2}{\Theta\omega^2}B_*\right) +
    \left(\tilde{C}+\frac{\omega_{p*}^2}{\Theta\omega^2}C_*\right) = 0.
\end{equation}
This is precisely the dispersion equation for the cold
plasma case except with $\omega_{p*}^2\rightarrow\omega_{p*}^2/\Theta$.
The resonance frequency in this case is
\begin{equation}
    \omega_{\infty}^{\left(1\right)}=\sqrt{\frac{\omega_{p*}^2}{\Theta}+\Omega^2},
\end{equation}
while the cutoff structure remains the same as the general case
discussed above.

\subsubsection{Low Frequency Asymptotic}
The $\omega\rightarrow0$ behavior
is given by taking the approximations
\begin{align}
\epsilon_{\perp}&\approx1+\frac{\omega_{p}^{2}}{\Omega^{2}}, \\
g_{*}&\approx\frac{\omega_{p}^{2}}{\omega\Omega}\frac{\Delta n}{n}.
\end{align}
In the cold plasma case,
\begin{equation}
    \epsilon_{\parallel,\text{cold}}=1+\alpha_{\epsilon}-\frac{\omega_{p*}^2}{\omega^2},
    \label{eq:epa-cold}
\end{equation}
which becomes large as $\omega\rightarrow0$ while $N^2$ remains finite.
Because of this, the $z$ polarization in Eq. (\ref{eq:disp-matrix})
is small compared to the other polarizations, 
so the low frequency behavior is determined
by the subspace
\begin{equation}
{\rm det}
\begin{bmatrix}
N^2\cos^2\theta -\epsilon_{\bot*} & -i\,g_*\\
i\,g_* & N^2\left(1-\alpha_\mu\sin^2\theta\right)-\epsilon_{\bot*}
\end{bmatrix} = 0.
\label{eq:disp-matrix-sub}
\end{equation}
In Case I,
\begin{equation}
    \epsilon_{\|*} = 1+\alpha_\epsilon-
    \frac{\omega_{p*}^2}{\Theta\omega^{2}\left(1-N^2\cos^{2}\theta\right)},
\end{equation}
which similarly becomes large as $\omega\rightarrow0$.
Therefore, the same subspace dominates and
the cold plasma behavior\cite{M23} is recovered in this limit.
For neutral plasma, this leads to dispersion relations
for the Alfv\'en and fast magentosonic waves
\begin{equation}
    N^2_+ = \frac{1+\omega_{p*}^2/\Omega^2}{\cos^2\theta},
\end{equation}
\begin{equation}
    N^2_- = \frac{1+\omega_{p*}^2/\Omega^2}{1-\alpha_\mu\sin^2\theta},
\end{equation}
while for non-neutral plasmas this leads to the
whistler wave disperson relation
\begin{equation}
    N^2 = \frac{1}{\cos\theta\left(1-\alpha_\mu\sin^2\theta\right)}
    \frac{\omega_{p*}^2}{\omega\Omega}
    \frac{\left|\Delta n\right|}{n}.
\end{equation}

\subsubsection{High Frequency Asymptotic}
\label{subsubsec:high-frequency}
For $\omega \rightarrow \infty$ case,
the plasma response is negligible,
amounting to taking $\omega_{p*}\rightarrow0$.
This case therefore reduces to the
vacuum, no plasma case which
is unchanged from the QED vacuum case \cite{M23}.
The vacuum dispersion relations are
\begin{equation}
    \omega = kc/N_{\perp,\parallel}
    \label{vac}
\end{equation}
with
\begin{equation}
    N_\perp^2=\frac{1}{1-\alpha_\mu \sin^2\theta},
    \label{nperp}
\end{equation}
\begin{equation}
    N_\parallel^2=\frac{1+\alpha_\epsilon}{1+\alpha_\epsilon \cos^2\theta}.
    \label{npara}
\end{equation}
The high frequency behavior is shared between
both Cases.

\subsection{Case II}
In Case II, the coefficients in Eq. (\ref{eq:N-cubic}) are
\begin{flalign}
 & \mathcal{A}=\tilde{A},\label{eq:caseii-begin}\\
 & \mathcal{B}=\tilde{B}+ \frac{2\Theta \omega_{p*}^2}{\omega^{2}\cos^{2}\theta} A_{*},\\
 & \mathcal{C}=\tilde{C}+ \frac{2\Theta \omega_{p*}^2}{\omega^{2}\cos^{2}\theta} B_{*},\\
 & \mathcal{D}= \frac{2\Theta \omega_{p*}^2}{\omega^{2}\cos^{2}\theta} C_{*}\label{eq:caseii-end}.
\end{flalign}
As before, general solutions can be found via the cubic formula
and are extremely algebraically complicated.
Case II only occupies the thin sliver of $\left(\omega,k\right)$ space
along $\omega\sim k_z$ (Fig. \ref{f:allow}),
so the general behavior is not typically important
especially in cases where $\Theta\gg1$.
In particular, cutoffs and perpendicular propagation
\textit{cannot} be considered in Case II due to
the singular behavior of $Q_*$
for $\omega\rightarrow$ and $\cos^2\theta\rightarrow 0$.
Below, we consider a few simple special cases for
the behavior of Case II.

\subsubsection{Parallel Propagation}
Under parallel propagation, $\theta=0$, 
there are three branches.
Two branches, given by Eq. (\ref{eq:parallel-shared}),
are shared with Case I.
The remaining branch,
corresponding to $\epsilon_{\parallel*}=0$, is
\begin{equation}
    N^{2}=\frac{2\Theta\omega_{p*}^{2}}{\omega^{2}\left(1+\alpha_{\epsilon}\right)}.
\end{equation}

\subsubsection{Neutral Plasma}
For a neutral plasma, $\Delta n/n=0$.
Then $g_*=0$ and Eq. (\ref{eq:disp-matrix}) immediately yields
three branches.
The first branch has dispersion relation
\begin{equation}
    N^{2}=\frac{\epsilon_{\perp*}}{1-\alpha_{\mu}\sin^{2}\theta}.
\end{equation}
The remaining branches are obtained from
\begin{equation}
    \left(N^{2}\cos^{2}\theta-\epsilon_{\perp*}\right)\left(N^{2}\sin^{2}\theta-\epsilon_{\parallel*}\right)-\left(N^{2}\sin\theta\cos\theta\right)^{2}=0.
\end{equation}
Expanding out $\epsilon_{\parallel*}$ and multiplying through by $N^2$
yields the bi-quadratic
\begin{equation}
    \mathsf{A}N^4+\mathsf{B}N^2+\mathsf{C} = 0
\end{equation}
with coefficients 
(not to be confused with the coefficients Eqs. (\ref{A}) - (\ref{C}))
\begin{flalign}
 & \mathsf{A}=\left(1+\alpha_{\epsilon}\right)\cos^{2}\theta+\epsilon_{\perp*}\sin^{2}\theta,\\
 & \mathsf{B}=\epsilon_{\perp*}\left(1+\alpha_{\epsilon}\right)+\frac{2\Theta\omega_{p*}^{2}}{\omega^{2}},\\
 & \mathsf{C}=\epsilon_{\perp*}\frac{2\Theta\omega_{p*}^{2}}{\omega^{2}\cos^{2}\theta}.
\end{flalign}
The remaining branches are obtained via the quadratic equation,
and are algebraically complicated.

\subsubsection{Low Frequency Asymptotic}
In the classical and cold plasma case, 
$\epsilon_{\parallel*}$ is given by Eq. (\ref{eq:epa-cold}),
which becomes large while $N^2$ remains finite,
leading to the reduction of the dispersion matrix to Eq. (\ref{eq:disp-matrix-sub}).
However, in Case II,
\begin{equation}
    \epsilon_{\|*} = 1+\alpha_\epsilon-
    \frac{2\Theta \omega_{p*}^2}{\omega^{2}N^2\cos^{2}\theta},
\end{equation}
which remains \textit{finite} as $\omega\rightarrow0$.
As such, in contrast to classical plasmas and Case I,
no such simple behavior can be obtained.

\subsubsection{High Frequency Asymptotic}
The high frequency behavior 
in Case II is identical to that of Case I,
which is discussed in Subsection \ref{subsubsec:high-frequency}.

\section{Numerical Results}
\label{s:num}

We present numerical solutions of the full dispersion
relation for variations of $B$, $\theta$, $\Delta n$, and $\Theta$.
\newtext{
In Figs. \ref{fig:branches}, \ref{fig:dispersion-compare-classical},
\ref{fig:dispersion-compare-QED},
we present the dispersion curves for selected choices of the parameters
to illustrate the differences between classical, cold, QED, and thermal
plasmas.
In Figs. \ref{fig:angle-b-grid-neutral} - \ref{fig:deln-temp-grid},
we perform full parameter sweeps to
demonstrate how the mode structure,
cutoff, and resonance frequencies
scale with the parameters.
}
In all figures,
dashed lines depict plasma eigenmode branches with
$N^2<0$, while solid lines depict propagating
modes with $N^2>0$.
The dispersion curves are for Case I only,
as the excluded area around $\omega\sim k_z$
is extremely thin for most values of $\Theta$
(see Fig. \ref{f:allow}).
For illustrative purpose, we chose the numerical values
of the plasma and cyclotron frequencies to be
$\omega_p=1$, $\Omega=3$.
We note that in a realistic magnetar magnetosphere, $\omega_p \ll \Omega$
by many orders of magnitude.
The units of $\omega$, $k$ are arbitrary, but
we set the speed of light $c=1$.

In all dispersion curve diagrams, the 
wave branches are labeled
as they would be for a classical plasma:
"A" is the Alfv\'en wave,
"F" is the fast magnetosonic wave,
"X" is the extraordinary oblique electromagnetic wave,
"O" is the ordinary oblique electromagnetic wave,
"W" is the whistler wave,
"Z" is the Z-mode,
and "L" is the Langmuir mode.

\subsection{Mode Identification}

Fig. \ref{fig:branches} helps us with identification of individual branches of the normal plasma modes. Such an identification is fairly straightforward in the case of a quasi-parallel  and a quasi-perpendicular propagation. 

In the quasi-parallel propagation case (top left panel), we have two modes at low frequencies, both have a linear dispersion, $\omega\propto k$. The upper one is the fast mode (F) and the lower one is the Alfv\'en (A) mode. Obviously, the Alfv\'en is slower at oblique angles because its phase speed depends on $k_\|$. Next, there is the longitudinal electrostatic Langmuir mode (L), starting at the plasma frequency cutoff (modified by temperature and QED effects) and extending to high frequencies with the linear dispersion proportional to $k_\|$. Being a longitudinal mode, it experiences no cyclotron resonance around $\omega\sim\Omega$, but it is generally heavily Landau damped. In contract, both electromagnetic modes propagating almost along the background field do experience cyclotron resonance in the pair plasma. The electromagnetic mode which has a component of its electric field along the background magnetic field is strongly affected by the QED-strength magnetic field. Namely, $N_\|>N_\bot$ in the QED regime, see Eqs. \eqref{nperp}, \eqref{npara}. This fact easily distinguishes the parallel polarization, which at large angles becomes the ordinary (O) mode. The second mode has the wave electric field orthogonal to the background field, so this polarization corresponds to the extraordinary (X) mode at large angles. The fast mode can be viewed as the lower-frequency (i.e., below the cyclotron resonance) extension of the X mode. 

In the quasi-perpendicular propagation case (bottom left panel), one can readily identify the O mode which experiences a sharp cyclotron resonance and which phase velocity strongly depends on the strength of the ambient magnetic field (compare solid orange and dashed blue curves). Similarly, we identify the X mode above the resonance and below the resonance, where it is labeled as the F mode. The Alfv\'en mode is clearly seen by its linear dispersion and low phase speed, whereas the Langmuir branch falls outside of the plotting box.

Identification of the modes at oblique angles (top right panel) may be somewhat perplexing. Still, the O, X and L modes above the cyclotron resonance are easily identifiable by their high-$\omega$, high-$k$ asymptotics. Similarly, the F (same as X) and A modes below the plasma frequency are well observed by their low-$\omega$, low-$k$ linear dispersion. At intermediate frequencies, one observes that the O mode is electromagnetically coupled to the non-quasi-neutral L mode. 

Finally, transitioning from the electrically neutral to the non-neutral plasma (bottom right panel), one observes that the low-frequency branches undergo modifications. The A mode exhibits a whistler-like quadratic dispersion, $\omega \propto k k_\|$, and is consequently labeled as W. The F mode transforms into the Z mode, characterized by a distinct new cutoff frequency that asymptotically approaches the plasma cutoff when $|\Delta n/n|$ approaches unity.

\subsection{Classical Plasma: Thermal Effects}
The index of refraction $N^2\left(\omega\right)$ and dispersion curves
$\omega\left(k\right)$ in a non-QED cold and thermal plasma
are shown in Fig. \ref{fig:dispersion-compare-classical}.
The cold case is plotted in green, while the
thermal case is plotted in blue.
The figure displays the mode structure for
cold and thermal plasma in a very weak field,
$B\ll B_Q$, for which quantum effects are negligible.
For illustrative purposes, we chose
the propagation angle $\theta=\pi/3$ and
numerical values of the plasma and cyclotron frequencies
to be $\omega_p=1$ and $\Omega=3$ respectively.
For the thermal plasma, we chose
temperature parameter $\Theta=10$.
This is for illustrative purposes,
as the ultrarelativistic regime
is for $\Theta \gg 1$.

Besides the obvious change
in mode structure
-- the propagation of the Alfv\'en mode
and the appearance of the Langmuir mode --
thermal effects are not very significant,
appearing only in the behavior near the
plasma frequency cutoff.

\subsection{QED Plasma: Dependence on magnetic field and propagation angle}
For both neutral and non-neutral plasma,
the QED effects are largely the same
between the cold and thermal plasma
cases.
This is most plainly seen in
Fig. \ref{fig:dispersion-compare-QED},
where we plot dispersion curves
for the cold and thermal
cases in the QED regime.
The cold case is plotted in purple,
and the thermal case is plottedin orange.
The dispersion curves
for most of the modes
lie on top of each other
between the cold and thermal cases.
The main differences appear
as $\omega\rightarrow0$,
where
$\omega_0^{\left(1\right)}$
is further reduced
from the cold plasma case.

\begin{figure*}
    \includegraphics[totalheight=0.3\paperheight]{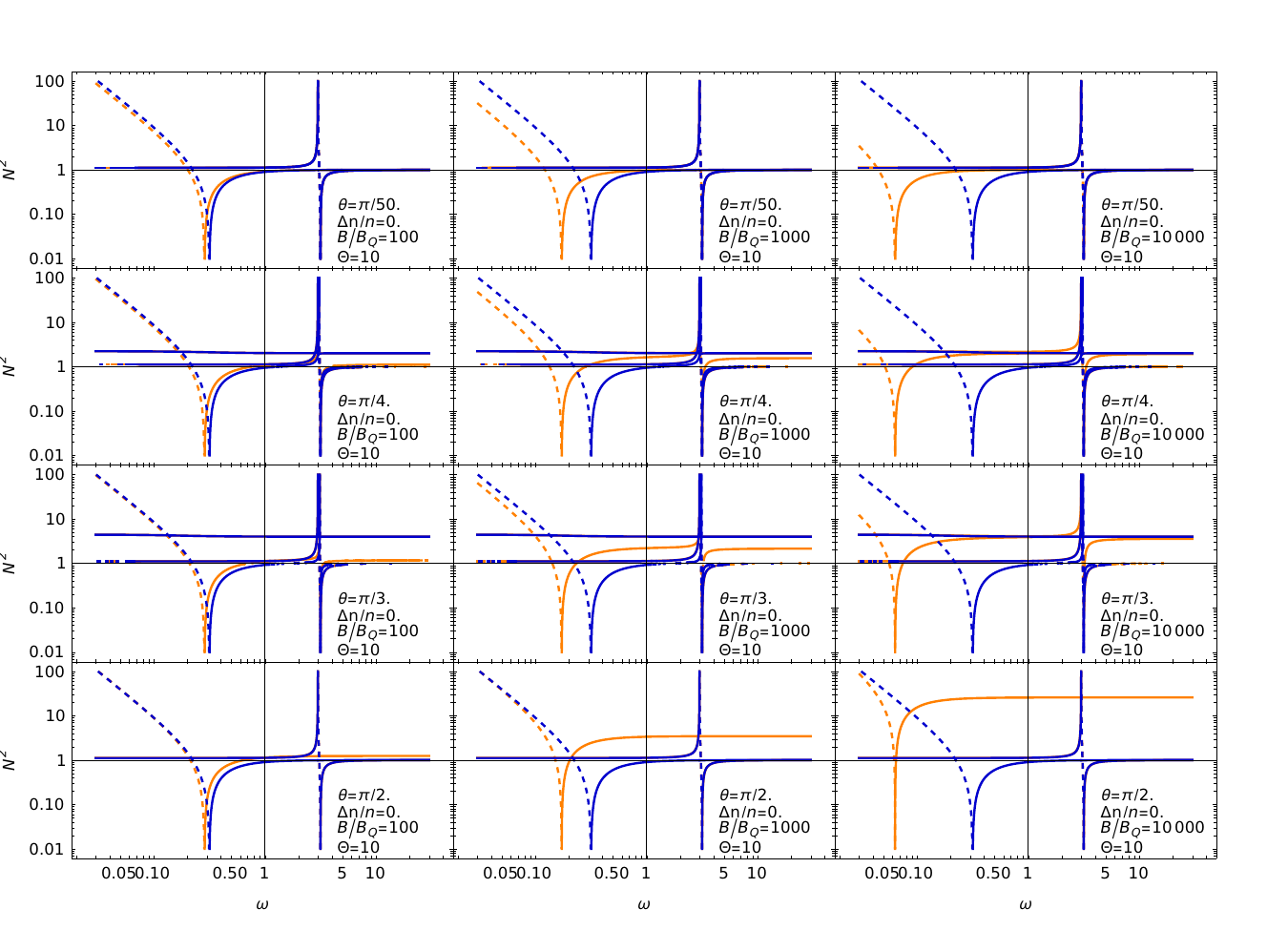}

    \includegraphics[totalheight=0.5\paperheight]{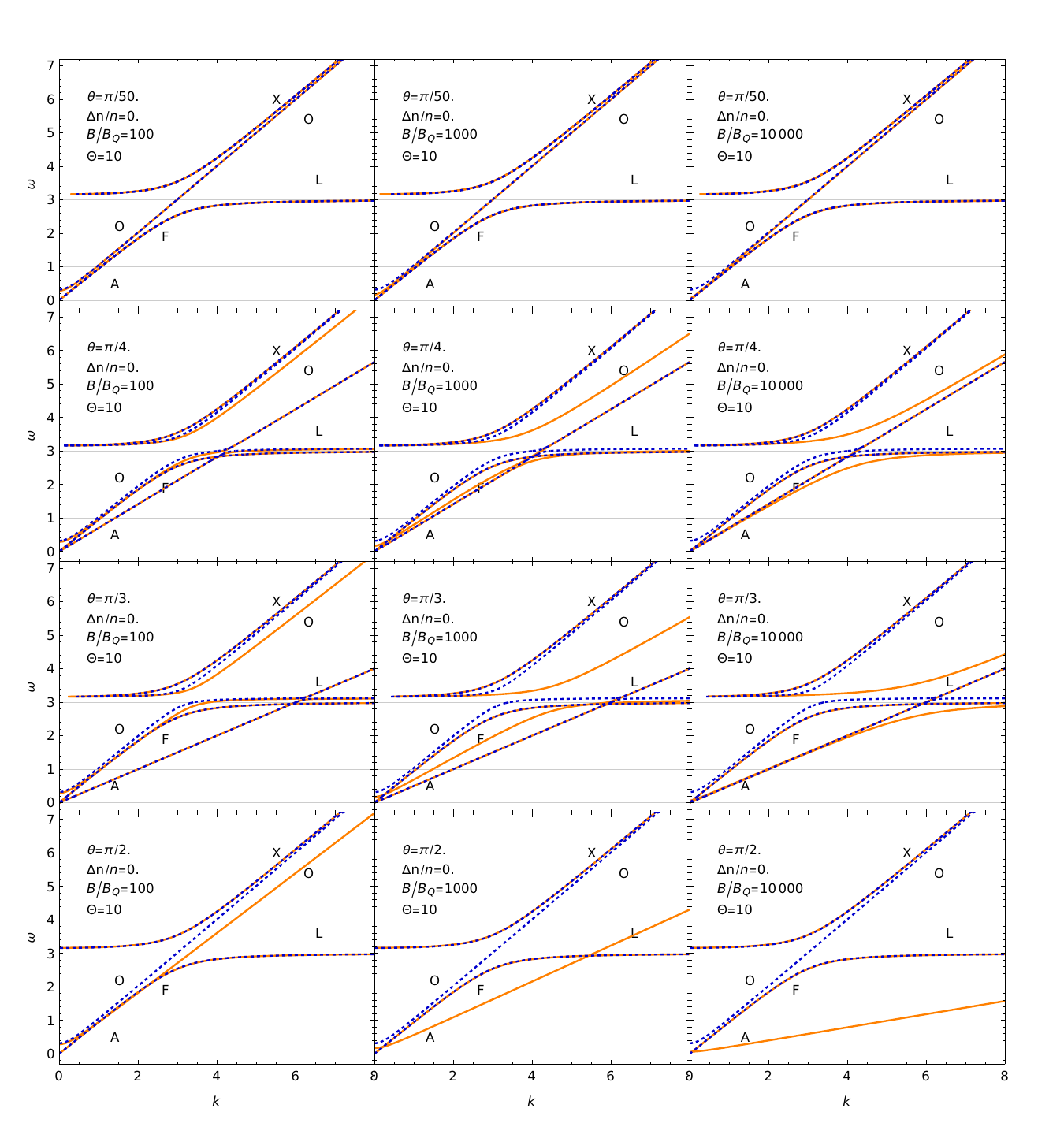}

    \caption{
    The index of refraction squared
    $N^2\left(\omega\right)$ (top row) 
    and the plasma dispersion curves 
    $\omega\left(k\right)$ (bottom row) 
    for the electrically neutral,
    $\Delta n/n=0$,
    QED plasma as functions
    of the magnetic field $B$,
    and the angle of propagation, $\theta$,
    with nearly parallel, oblique
    and perpendicular propagation.
    The mode structure for
    perpendicular propagation, $\theta=\pi/2$,
    is qualitatively different from the general case
    since it reduces to the cold plasma
    case \cite{M23}.
    The blue curves illustrate the non-QED regime and are
    shown for comparison.
    The temperature parameter is $\Theta=10$. 
    The plasma frequency is
    $\omega_p=1$ and the cyclotron frequency is $\Omega=3$.
    The latter is set to a constant, despite varying $B$, for the ease of comparison.
    The wave branches are labeled as in Fig. \ref{fig:dispersion-compare-classical}.
    }
    \label{fig:angle-b-grid-neutral}
\end{figure*}

\begin{figure*}
    \includegraphics[totalheight=0.3\paperheight]{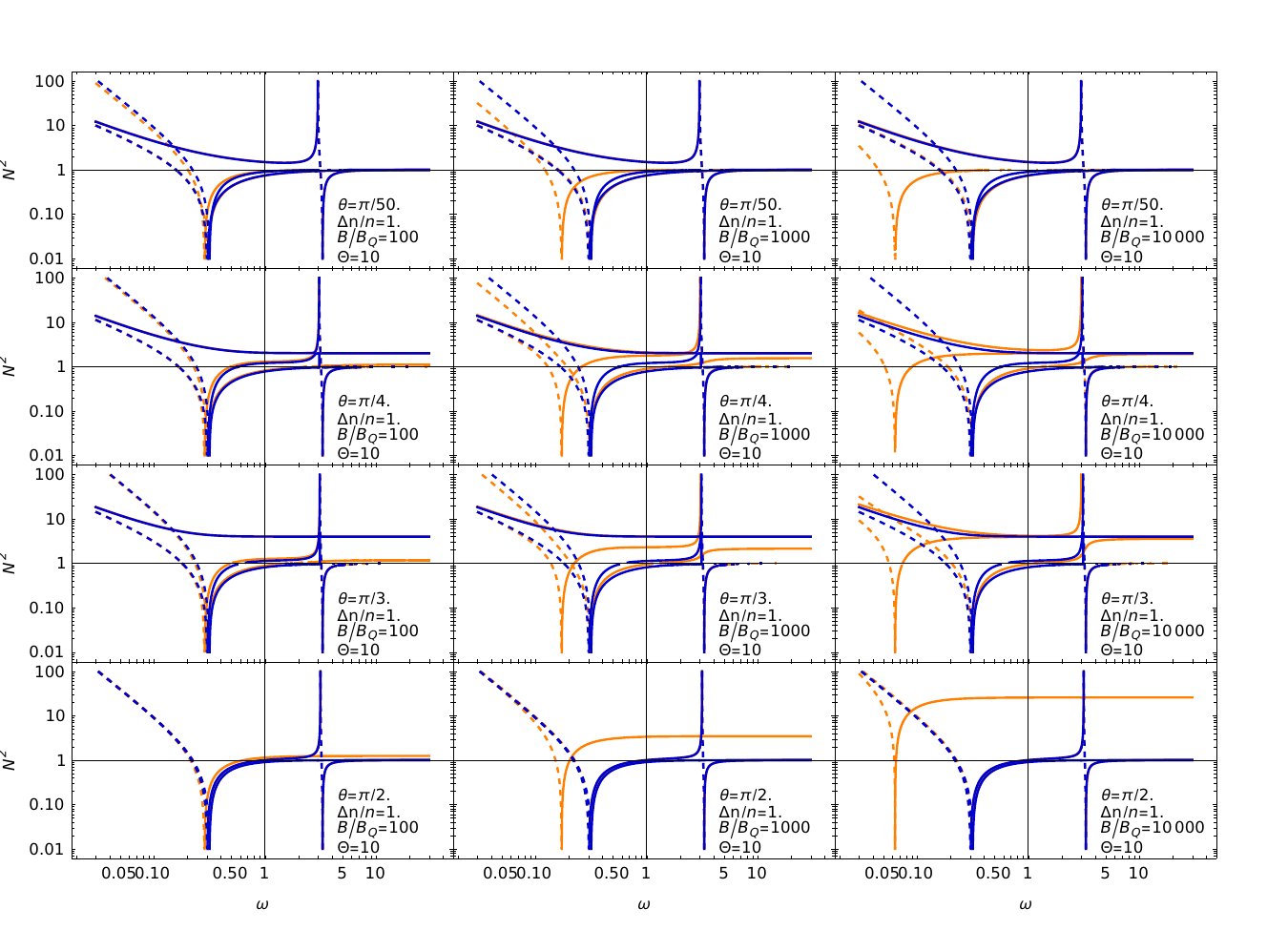}

    \includegraphics[totalheight=0.5\paperheight]{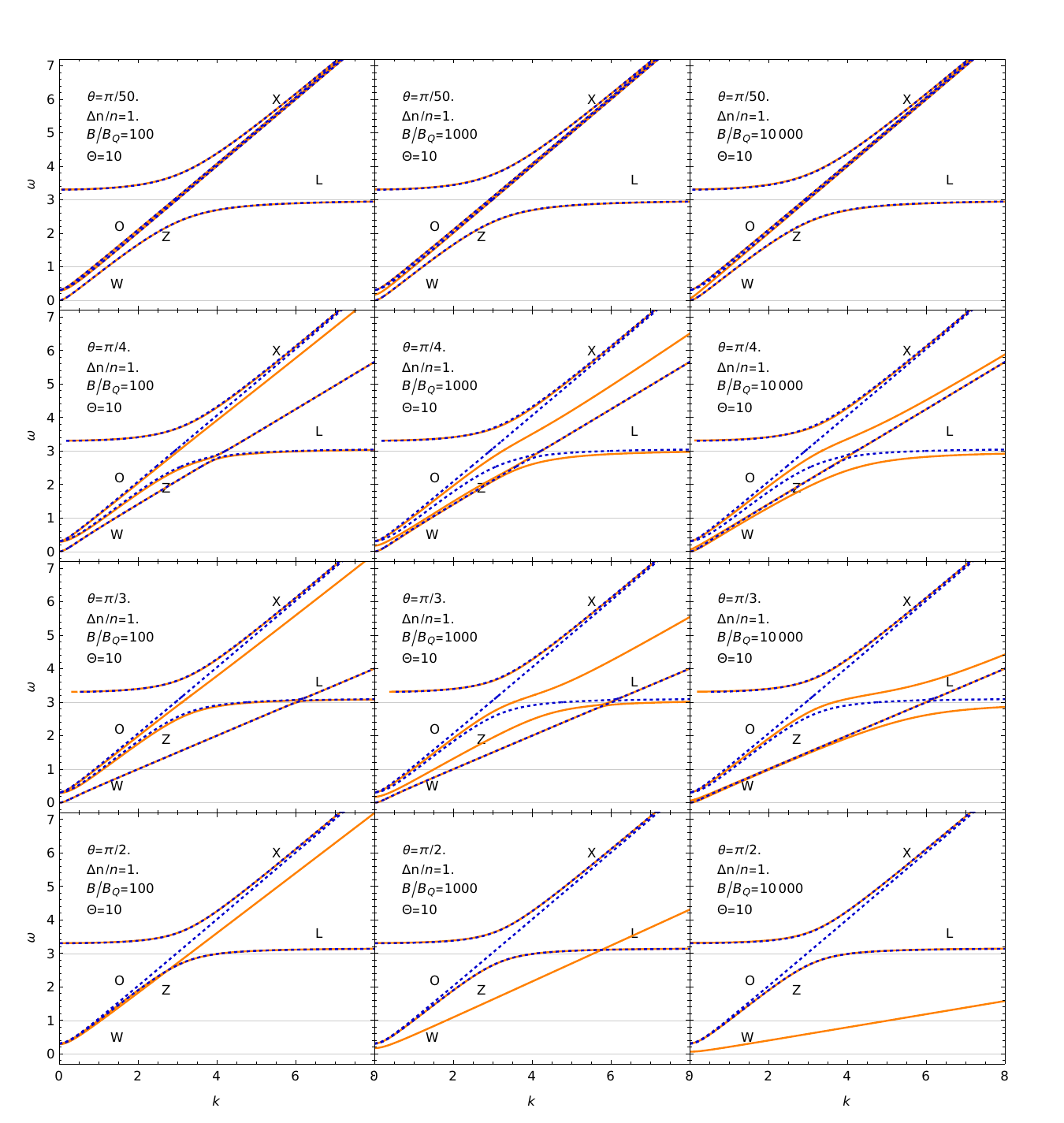}

    \caption{
    The index of refraction squared
    $N^2\left(\omega\right)$ (top row) 
    and the plasma dispersion curves 
    $\omega\left(k\right)$ (bottom row) 
    for the electrically non-neutral,
    $\Delta n/n=1$,
    QED plasma as functions
    of the magnetic field $B$,
    and the angle of propagation, $\theta$,
    with nearly parallel, oblique
    and perpendicular propagation.
    The mode structure for
    perpendicular propagation, $\theta=\pi/2$,
    is qualitatively different from the general case
    since it reduces to the cold plasma
    case \cite{M23}.
    The blue curves illustrate the non-QED regime and are
    shown for comparison.
    The temperature parameter is $\Theta=10$. 
    The plasma frequency is
    $\omega_p=1$ and the cyclotron frequency is $\Omega=3$.
    The latter is set to a constant, despite varying $B$, for the ease of comparison.
    The wave branches are labeled as in Fig. \ref{fig:dispersion-compare-classical}.
    }
    \label{fig:angle-b-grid-not}
\end{figure*}

In Fig. \ref{fig:angle-b-grid-neutral}, we present
the dispersion curves (solid orange lines) for
an electrically neutral plasma, $\Delta n=0$, in
superstrong $B$-field of strengths
$10^2\leq B/B_Q \leq 10^4$
for various angles of propagation.
The dispersion curves for a non-QED plasma,
that is with $B/B_Q\rightarrow0$,
are shown in blue dashed curves for comparison.
Similar results are presented for
non-neutral plasma in Fig. \ref{fig:angle-b-grid-not}.

For perpendicular propagation,
the system reduces to the
cold plasma behavior except
with $\omega_{p*}^2\rightarrow\omega_{p*}^2/\Theta$.
In particular, the O-mode has no cyclotron
resonance and is appreciably slowed
by QED effects
while the X- and fast modes
are not appreciably affected.

The main QED effect,
much like in the cold plasma
case,
is the slowing and angle
dependence of the O-mode.

\subsection{QED Plasma: Dependence on temperature and non-neutrality}
\begin{figure*}
    \includegraphics[totalheight=0.3\paperheight]{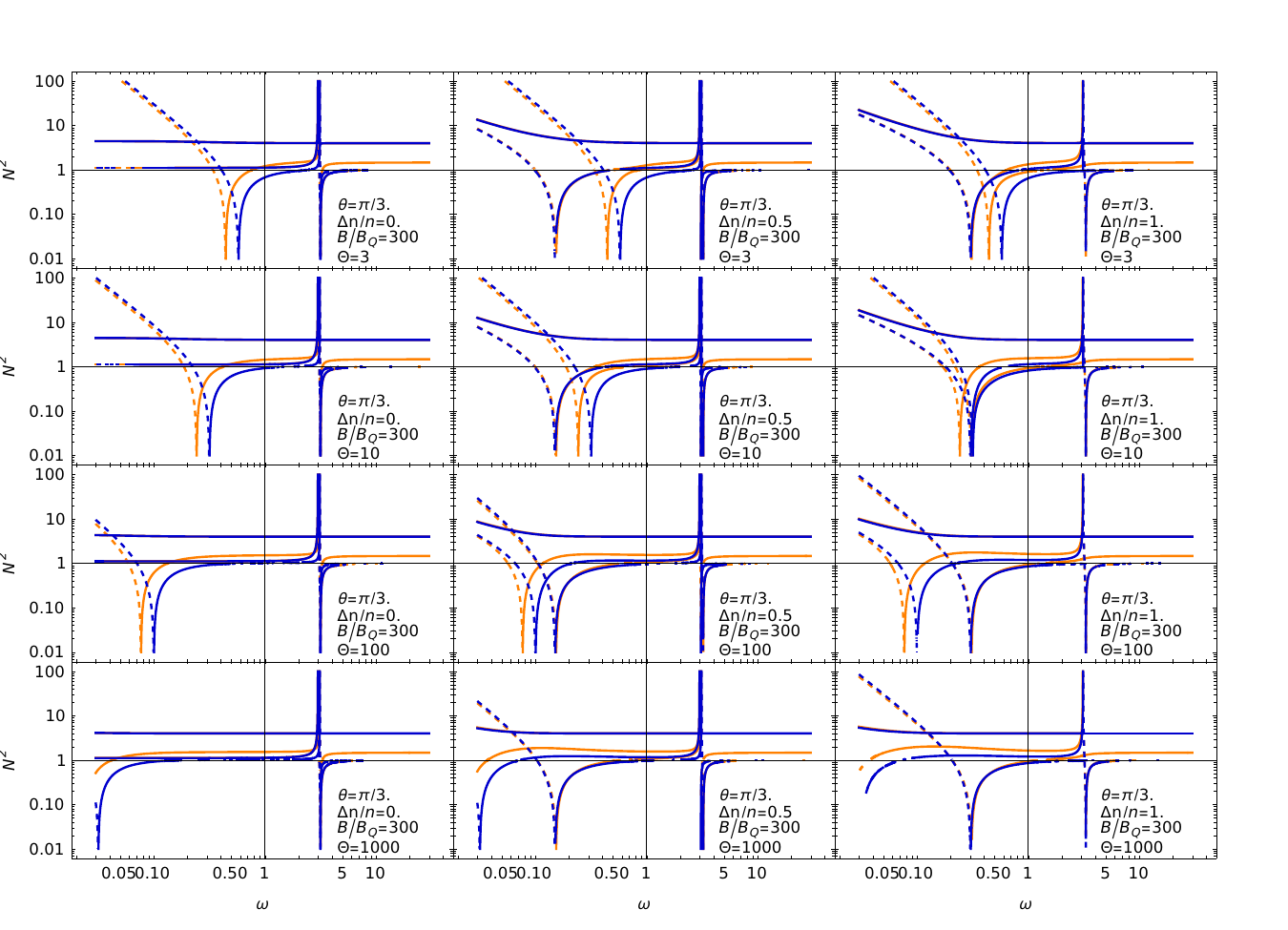}

    \includegraphics[totalheight=0.5\paperheight]{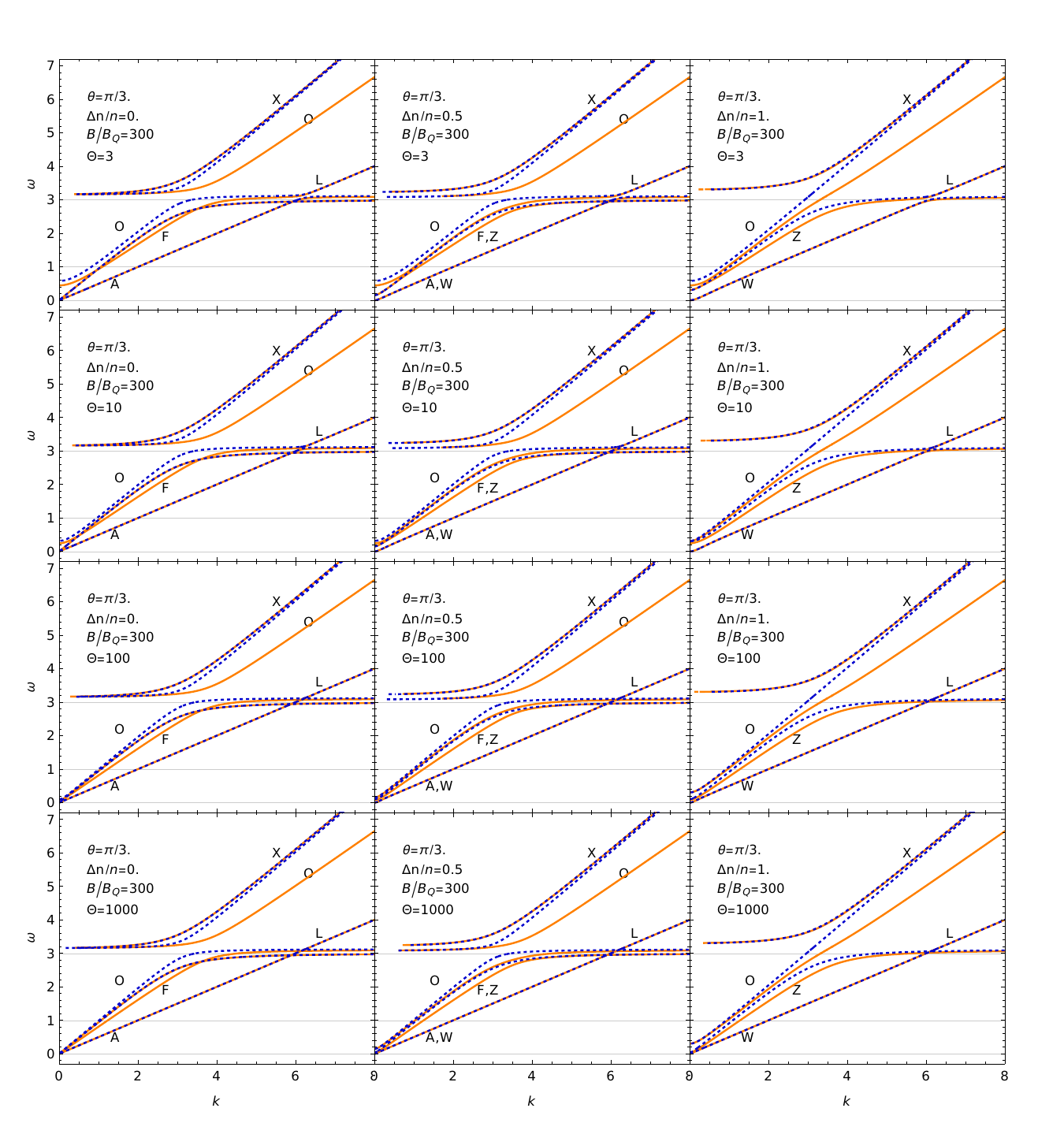}

    \caption{
    The index of refraction squared
    $N^2\left(\omega\right)$ (top row) 
    and the plasma dispersion curves 
    $\omega\left(k\right)$ (bottom row) 
    as a function of the non-neutrality parameter,
    $\Delta n/n$, and the temperature parameter $\Theta$.
    The $\Theta=3$ case is for illustrative
    purposes only since it is not within the
    ultrarelativistic limit $\Theta\gg1$.
    The plasma is in the QED regime with 
    $B/B_Q=300$ and $\theta=\pi/3$.
    The blue curves illustrate the non-QED regime and are
    shown for comparison.
    The plasma frequency is
    $\omega_p=1$ and the cyclotron frequency is $\Omega=3$.
    The latter is set to a constant, despite varying $B$, for the ease of comparison.
    The wave branches are labeled as in Fig. \ref{fig:dispersion-compare-classical}.
    }
    \label{fig:deln-temp-grid}
\end{figure*}

In Fig. \ref{fig:deln-temp-grid}, we present
the dispersion curves for $B=300B_Q$, $\theta=\pi/3$
while we vary $\Delta n=0, 0.5, 1$ and
$\Theta=3, 10, 100, 1000$.
The case $\Theta=3$ is for illustrative purposes
only as it does not lie within the ultrarelativistic regime, $\Theta \gg 1$.
As discussed in Sec. \ref{s:analysis}, 
the cutoff frequency $\omega_0^{\left(1\right)}$
scales with temperature as $\Theta^{-1/2}$,
so the main effect of increasing $\Theta$
is $\omega_0^{\left(1\right)}\rightarrow0$.
The mode structure otherwise remains relatively
unchanged by increasing $\Theta$.

Non-neutrality changes the mode structure
similarly to the cold plasma case:
the O-mode becomes non-resonant near
the cyclotron frequency as
$\left|\Delta n/n\right|\rightarrow 1$
while the associated cutoff, $\omega_0^{\left(1\right)}$,
is unaffected by non-neutrality.

\begin{figure*}
    \includegraphics[width=0.8\textwidth]{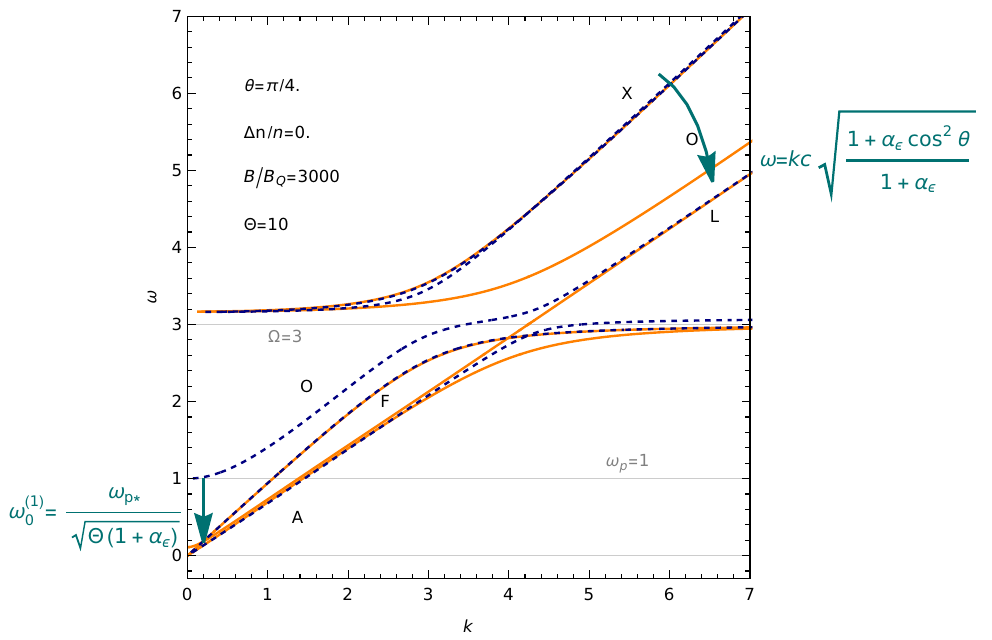}
    \caption{
    Summary figure of plasma dispersion curves $\omega\left(k\right)$, showing
    QED effects on charge-neutral, $\Delta n=0$, thermal plasma.
    We set $B=3000B_0\sim10^{17} \text{ G}$, $\theta=\pi/4$, and
    $\Theta=k_B T/\left(m_e c^2\right)=10$.
    The plasma frequency is $\omega_p=1$ and the cyclotron frequency
    is $\Omega=3$.
    For comparison,
    dashed blue curves represent the dispersion curves in a non-QED plasma
    with minimal thermal effects,
    i.e. for $B/B_Q\rightarrow0$ and $\Theta=1$.
    The wave branches are labeled as follows: 
    “A”—Alfv\'en wave, 
    “F”—fast magnetosonic wave, 
    “X”—extraordinary electromagnetic wave, 
    “O”—ordinary electromagnetic wave 
    (in a neutral plasma, it consists of two branches split around the cyclotron frequency),
    "L"-Langmuir mode.
    The most pronounced QED effects are shown by arrows.
    They are:
    (i) the $B$-field induced transparency of the O-mode
    seen in the reduced wave cutoff frequency, $\omega_0^{\left(1\right)}$
    as $k\rightarrow0$ and
    (ii) the reduction of the phase speed of the O-mode at $\omega\gg\omega_{p*}$.
    Thermal effects are also present.
    Resonant and cutoff frequencies typically scale as $\Theta^{-1/2}$,
    in this case driving $\omega_0^{\left(1\right)}$ even further towards zero
    and the other resonances towards the cyclotron resonance.
    }
    \label{fig:summary-fig}
\end{figure*}

\section{Summary}
\label{s:summary}

Utilizing the QED-plasma framework,
we have derived dispersion relations of normal modes in a non-neutral ultrarelativistic pair plasma that is embedded in a QED-strong background magnetic field.
\newtext{
These results extend the cold plasma case explored in Paper I
to a plasma with relativistic temperature effects.
Many results from the cold QED plasma are recovered
in the thermal plasma case.
}
We obtained the following results, summarized in Fig. \ref{fig:summary-fig}.
\begin{enumerate}
    \item Many effects from the cold plasma presented in Paper I \cite{M23}
        are retained. In particular the
        retention of the classical plasma mode structure
        (no novel, QED-only modes appearing),
        the renormalization
        of the plasma frequency (see Eq. \ref{alphaem},
        Fig. \ref{f:waa})
        \begin{equation}
            \omega_{p*}=\frac{\omega_p}{\left(1-C_\delta\right)},
        \end{equation}
        and the entrance of QED corrections
        via $\alpha_\epsilon$ and $\alpha_\mu$. \newtext{Note that the modification of the plasma frequency in an ultramagnetized QED-plasma is independent of the thermal effects and is associated with global energy density modifying the isotropic susceptibility of vacuum.}
    \item In a QED plasma,
        increasing $B$-field strength
        allows
        the O-mode
        to propagate at frequencies below
        the plasma frequency.
        Thermal effects further enhance
        this effect, as seen in Eq. (\ref{eq:cutoff-generalize}),
        \begin{equation}
            \omega_{0}^{\left(1\right)}=\frac{\omega_{p*}}{\sqrt{\Theta\left(1+\alpha_{\epsilon}\right)}}.
        \end{equation}
        \newtext{This effect is very interesting. One can see that the O-mode cut-off is lowered by both the ultra-strong magnetic field and the thermal effect. The former is induced by the anisotropic contribution to dielectric susceptibility (i.e., permittivity) of vacuum via the coefficient $\alpha_{\epsilon}$, whereas the contribution to the magnetic permeability is negligible. The latter (thermal) modification enters as $\omega_{p}/\sqrt{\Theta}$. Noting that $\Theta$ is essentially the thermal Lorentz factor of the plasma, we conclude that the effect is similar to what is known as ``relativistic transparency''. }
    \item Similar to the cold plasma case,
    the ordinary mode is slowed
    as seen in the increase of the
    index of refraction, Eq. \ref{npara}.
    At high frequencies, $\omega \gg \omega_p$
    it has the QED vacuum dispersion relation
    \begin{equation}
        \omega=kc\sqrt{\frac{1+\alpha_\epsilon \cos^2\theta}{1+\alpha_\epsilon}}.
    \end{equation}
    \newtext{We emphasize that this effect appears to be independent of the plasma temperature, so it is the same in cold and relativistic plasmas.}
\end{enumerate}
We summarize our results in Fig. \ref{fig:summary-fig}.
We plot the plasma dispersion curves (solid orange) for a charge-neutral,
thermal QED plasma with $B\sim10^{17} \text{ G}$ for
oblique propagation, $\theta=\pi/4$.
This is compared to the
non-QED plasma modes (dashed blue) with minimal thermal
effects, i.e. with $B/B_Q\rightarrow0$ and $\Theta=1$.
Labeled are the
Alfv\'en (A),
fast magnetosonic (F),
Langmuir (L),
ordinary (O),
and extraordinary (X)
modes.
We set the plasma frequency $\omega_p=1$
and cyclotron frequency $\Omega=3$.

These results are of particular importance for neutron star and magnetar magnetospheres. They would help one better understand radiation propagation through these environments, including the origin of FRBs, better constrain nuclear equation of state via more accurate X-ray hot-spot reconstruction in pulsars, and more.

\begin{acknowledgments}
This research was supported by the National Science Foundation under Grant PHY-2409249. 
\end{acknowledgments}

\bibliography{references}{}
%\bibliographystyle{aasjournal}



\end{document}